\documentclass[9pt]{elife}
\newcommand{\fig}{Fig.~}
\usepackage[section]{placeins}
\usepackage{lipsum} 
\usepackage[version=4]{mhchem}
\usepackage{siunitx}
\DeclareSIUnit\Molar{M}

\title{Reassessing prediction in the brain: Pre-onset neural encoding during natural listening does not reflect pre-activation }

\author[1,2*]{Sahel Azizpour}
\author[1,3\authfn{1}]{Britta U. Westner}
\author[4\authfn{1}]{Jakub Szewczyk}
\author[1,2]{Umut Güçlü}
\author[1,2]{Linda Geerligs}
\affil[1]{Donders Institute for Brain, Cognition, and Behaviour, Nijmegen, Netherlands}
\affil[2]{Radboud University, Nijmegen, Netherlands}
\affil[3]{Radboud University Medical Center, Nijmegen, Netherlands}
\affil[4]{Jagiellonian University, Kraków, Poland}

\corr{sahel.azizpour@donders.ru.nl}{}

\contrib[\authfn{1}]{These authors contributed equally to this work}

\begin{document}

\maketitle

\begin{abstract}
Predictive processing theories propose that the brain continuously anticipates upcoming input. However, direct neural evidence for predictive pre-activation during natural language comprehension remains limited and debated. Previous studies using large language model (LLM)-based encoding models with fMRI and ECoG have reported pre-onset signals that appear to encode upcoming words, but these effects may instead reflect dependencies in the stimulus or autocorrelations in neural activity.

Here, we re-examined this question by aligning LLM-derived word embeddings with neural activity recorded during naturalistic listening using magnetoencephalography (MEG) and electrocorticography (ECoG). We replicated pre-onset encoding effects previously observed in ECoG across both modalities, and found that they persist even after controlling for stimulus correlations. Crucially, temporal generalization analyses revealed no stable overlap between pre- and post-onset representations, indicating that pre-onset activity does not reflect pre-activation of the next word. Consistent with this, long-range predictive effects previously reported in fMRI did not replicate in our higher-temporal-resolution data.

While we found no evidence for predictive pre-activation, we observed clear signatures of postdiction, with neural activity reflecting persistent encoding of prior words. These results suggest that reported apparent predictive signals do not reflect pre-activation of upcoming input. They call for caution in interpreting LLM-based encoding models and highlight the need for a more nuanced understanding of what constitutes “prediction” in language comprehension.
\end{abstract}

\section{Introduction}

Predictive processing is a foundational hypothesis in neuroscience, that proposes that the brain is continuously engaged in making predictions about upcoming stimuli \citep{clark2013whatever, friston2010free}. These predictions are shaped by the context in which the input is presented and are informed by the individual's previous experiences with similar input sequences \citep{de2018expectations}. Language processing is one domain where predictions are thought to play a crucial role, particularly in facilitating comprehension by pre-activating linguistic representations \citep{federmeier2007thinking, kutas2011look}. Various approaches have been used to investigate how the availability of predictive information influences lexico-semantic processing. At the behavioral level, studies have shown that highly predictable words result in shorter reaction times and reading times \citep{smith2013effect, traxler2000effects}. At the neural level, numerous studies have demonstrated that the ERP response is modulated by the predictability of a word within a contextually constraining sentence \citep{delong2005probabilistic, kutas2011look}. Compelling evidence from studies of natural conversations also shows that the absence of temporal gaps between exchanges requires listeners to predict the upcoming word and content with high accuracy \citep{ de2006projecting, magyari2012prediction}. Over the past five decades, a wealth of research has strongly suggested that language comprehension is inherently predictive \citep{kuperberg2016we}.

Despite broad agreement that language comprehension involves prediction, there is less agreement on what form these predictions take and how they influence processing. Some accounts emphasize pre-activation, in which features of upcoming words (e.g., semantic, syntactic, or phonological) are actively represented before they appear \citep{federmeier2007thinking,antonello2024predictive}. Others argue that any facilitation in word processing necessarily implies some form of prediction \citep{kuperberg2016we}. However, such predictions may reflect a representation of the word already integrated with the context, rather than pre-activation of its features. Empirically disentangling these accounts remains challenging, as traditional neural measures such as the N400 can reflect both pre-activation of word-related features and post-hoc processing of input \citep{kutas2011thirty}. This ongoing debate highlights the need for methods that can directly test pre-activation accounts of prediction by capturing the representational content of prediction during continuous language comprehension.

Evidence for predictive pre-activation was provided by \citep{wang2020neural, wang2024dissociating}. They used representational similarity analysis (RSA) to show that EEG patterns just before a highly predictable target word could reveal whether the noun was animate. However, their approach required a one-second gap before the target. While this interval presumably improved the ability to decode the noun's animacy, it may have encouraged participants to adopt strategic, unnatural predictive processes that are less likely in natural comprehension scenarios.

Recent advancements have created new opportunities to study predictive processing in naturalistic settings. High-quality datasets from MEG and ECoG now allow researchers to examine prediction across tens of thousands of words, such as in participants listening to entire audiobooks. These large-scale datasets offer greater statistical power to detect subtle predictive representations while exposing participants to richer linguistic contexts with higher ecological validity than traditional experimental designs \citep{hamilton2020revolution, willems2020narratives}. Crucially, in such conditions, a word’s predictability is shaped by long-range discourse rather than just the preceding word or sentence.

A potential challenge with using large datasets has been quantifying word predictability, as traditional cloze probability tests are labor-intensive and infeasible for datasets with tens of thousands of words. Fortunately, recent advances in LLMs overcome this limitation. Trained on vast text corpora, LLMs capture nuanced linguistic representations and can serve as proper linguistic theories \citep{baroni2022proper}, and by extension, proxies for the brain’s language processes. Indeed, across various neuroimaging modalities, such as fMRI \citep{caucheteux2022brains, caucheteux2023evidence,schrimpf2021neural}, MEG \citep{caucheteux2022brains, toneva2022combining}, and ECoG \citep{goldstein2022}, it has been demonstrated that LLM-generated word embeddings can be linearly mapped onto brain activity, explaining a substantial portion of the variance in neural responses to language.

Leveraging the power of LLMs, only a few studies have examined neural signatures of predictive pre-activation during naturalistic narrative listening. The first, by \citet{goldstein2022}, examined whether neural signals preceding word onset contain predictive information about upcoming words. Using high-precision ECoG recordings during natural speech listening, they found that the brain represents the upcoming word up to two seconds before the word onset. This finding has since been replicated and extended to examine various properties of predictive processing in natural language comprehension \citep{goldstein2022correspondence, zada2024shared}. 

A second line of evidence comes from \citet{caucheteux2023evidence}, who demonstrated that fMRI responses during natural story listening were better predicted when models incorporated embeddings of both current and future words. This result was interpreted as reflecting the brain’s ability to represent multiple upcoming words, providing potential evidence for long-range predictive processing.

Both studies suggest that the brain actively predicts upcoming linguistic input before it is perceived. However, their interpretations remain debated. \citet{schonmann2025stimulus} cautioned that the pre-onset signals reported by \citet{goldstein2022} may not reflect genuine pre-activation of lexical representations. They showed that similar pre-onset patterns can emerge when models predict the acoustics of continuous speech rather than neural activity. This finding indicates that apparent pre-onset encoding may partly arise from correlations inherent in the stimulus, such as dependencies between adjacent word embeddings or co-occurring acoustic features, rather than from predictive neural processes.

Similarly, \citet{caucheteux2023evidence} relied on fMRI data with inherently low temporal resolution, averaging word embeddings across 1.5 s repetition times (TRs). This temporal smoothing could obscure word-level dynamics and introduce autocorrelations that inflate apparent prediction effects.

Here, we revisit these claims using a multimodal approach to test (1) whether previously reported pre-activation signals can be replicated across different datasets and modalities, and (2) whether such effects truly reflect predictive pre-activation rather than other sources of temporal correlation.

To this end, we mapped LLM-derived word embeddings onto neural activity from two complementary datasets: (1) broadband MEG recordings collected while participants listened to approximately 10 hours of narrative speech, and (2) high-gamma ECoG signals recorded as participants listened to ~30 minutes of a podcast (the same dataset as reported in \citet{goldstein2022}).

Across both modalities, we replicated pre-onset encoding effects reported in prior work \citep{goldstein2022} and confirmed that these effects persist even after controlling for statistical regularities in the stimulus, such as correlations among adjacent word embeddings. However, when applying Temporal Generalization (TG) analysis, we found no evidence for stable representational overlap between pre- and post-onset time points. This suggests that pre-onset encoding does not reflect the active pre-activation of specific upcoming word representations.

In line with this finding, when adapting the fMRI analysis of \citet{caucheteux2023evidence} to our MEG and ECoG data, we found no improvement in neural encoding when including embeddings of future words. Instead, we observed robust evidence for postdiction: incorporating embeddings of preceding words improved model performance, indicating that prior word representations persist after word offset.

Our results suggest that previously reported predictive signals do not reflect genuine pre-activation of upcoming input. Instead, they call for caution when using LLM-based encoding models to study predictive processing and highlight the need for a more nuanced discussion of what constitutes “prediction” in language comprehension, and whether prediction in this context truly entails pre-activation.

\section{Results}

\begin{figure*}
    \hspace{-0cm} 
    \centering
    \includegraphics[scale=1,width=1\textwidth]{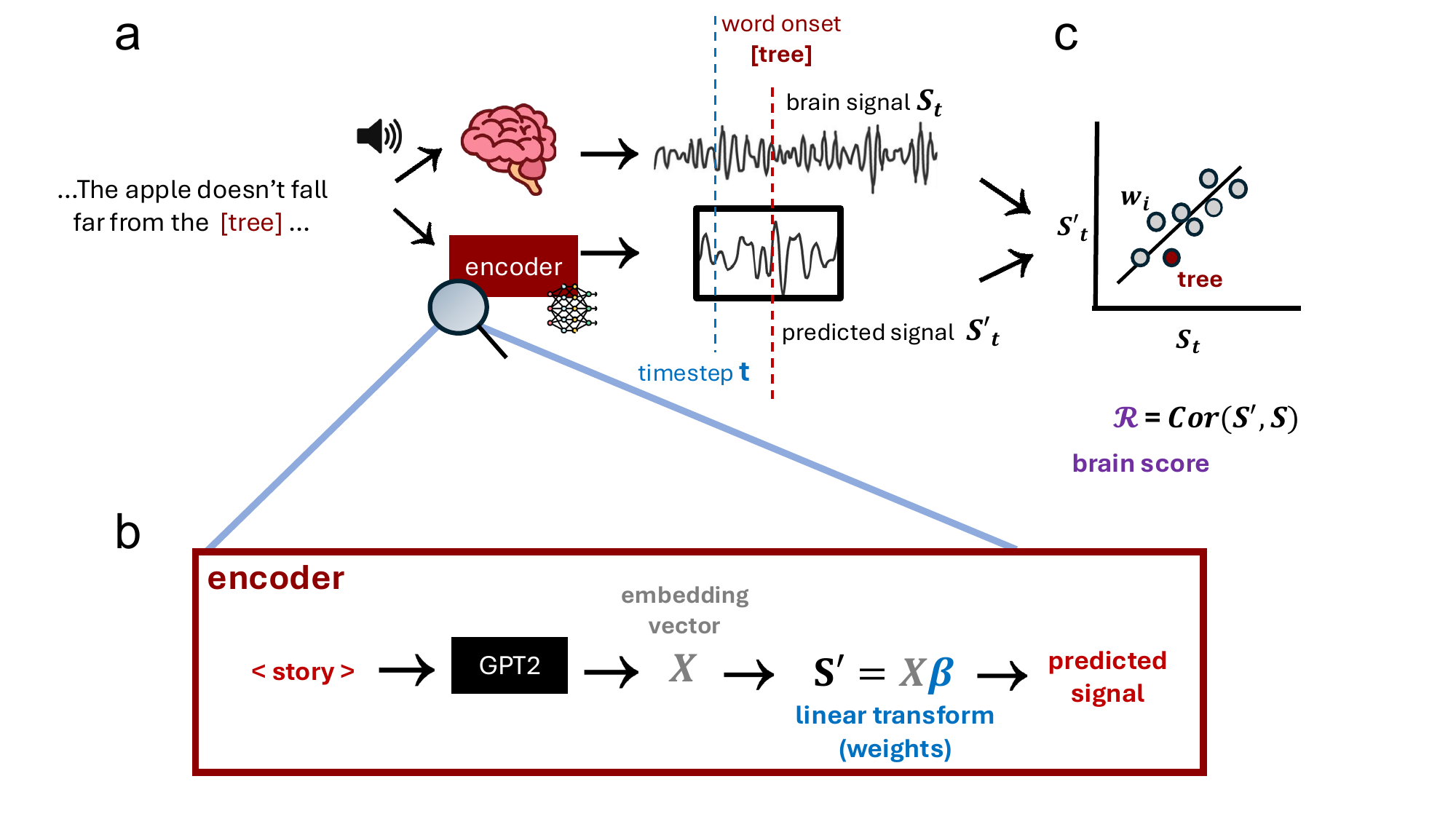}
    \caption{\textbf{The encoding model finds a linear mapping between words in the narrative and corresponding brain responses.}
    \textbf{a}. The encoding model mimics the brain by taking each word and its context and learning to generate a brain-like response within a time window around the word’s onset.
    \textbf{b.} Word embeddings are extracted from the GPT-2 model and used to predict brain responses at each time point $t$ through linear regression.
    \textbf{c.} The brain score represents the correlation between actual and predicted brain responses across multiple words, calculated at each time point $t$.}
    \label{fig:encoding_scheme}
\end{figure*}

The encoder processed the words from the narrative and predicted corresponding brain signals within a time window around word onset (\fig\ref{fig:encoding_scheme}). It was trained to learn a linear mapping between the LLM's activations (word embeddings) and the MEG broadband signal (0.1-40 Hz) and the ECoG high gamma (70-200 Hz) broadband power recorded at each time point within this window. To quantify how well the LLM activations capture information in the brain activity, we computed the correlation between the predicted and actual brain responses. The resulting correlation at each time point, termed the 'brain score' $\mathcal{R} $, indicates the extent to which the word's features are represented in the brain signal. 

\begin{figure}[htbp]
    \centering
    \includegraphics[scale=1,width=1\textwidth]{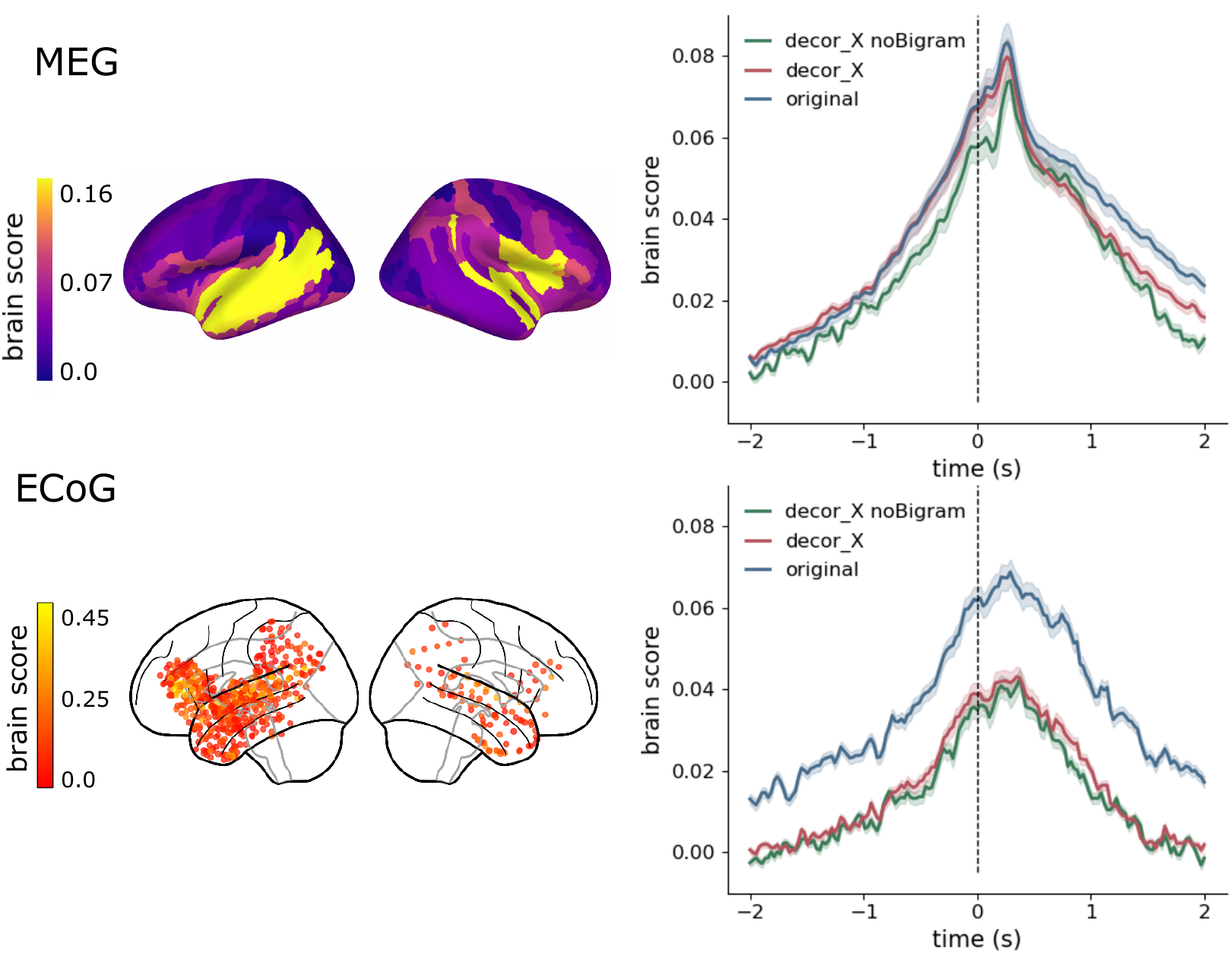}
    \caption{\textbf{Word embeddings explain brain responses before word onset}
    \textbf{left column:} All brain regions show encoding of the word embedding, with peak values in the left hemisphere, especially in the temporal cortex, inferior frontal areas and tempoparietal junction associated with language processing. \textbf{right column:} A comparison of encoding models shows that removing correlations between neighboring word embeddings (decor\_X) does not remove pre-onset encoding. Similarly, eliminating bi-grams in the narrative (decor\_X noBigram) slightly reduces overall encoding performance, but the pre-onset effect persists.
    The shaded regions in the line plots indicate the standard error of the mean (SEM) across aggregated ECoG electrodes/MEG sources of all participants.}
    \label{fig:fig2}

\end{figure}

Encoding models were trained separately for each participant, time point, and MEG source or ECoG electrode. In MEG, nearly all regions exhibited positive correlations between predicted and observed brain responses, with the strongest effects in the left temporal gyri and inferior frontal regions (\fig\ref{fig:fig2}.a). Similarly, all ECoG electrodes showed positive encoding of word embeddings, with maximum correlation values larger than MEG. The strongest effects were observed in the left inferior frontal gyrus and the temporo-parietal junction, areas classically associated with language processing (\fig\ref{fig:fig2}.a).  To examine the temporal dynamics of the brain score around word onset, we first computed average brain scores across a predefined set of MEG sources and ECoG electrodes (see Methods) and then averaged within each set. As shown in \fig\ref{fig:fig2}.b, encoding profiles peaked approximately 300 ms after word onset, with ECoG showing lower average values given the broad electrode selection compared to MEG. A session-by-session analysis ($\sim$\,1 hour per session) of individual participants in the MEG dataset revealed stable encoding across sessions, with brain scores comparable to those obtained when training on the full 10-hour dataset (\fig\ref{fig:supp_MEG_fig2_per_subj}). In contrast, analyses of individual participants in the ECoG dataset showed greater variability across participants (\fig\ref{fig:supp_ECoG_fig2_per_subj}), consistent with differences in electrode coverage across subjects and the shorter session length (~30 min). 

\subsection{Word embeddings explain brain responses before word onset}

Notably, positive brain scores are observed up to one second before the word onset, indicating consistent pre-onset neural representations across modalities (\fig\ref{fig:fig2}.b),  with these pre-onset effects being stronger for predictable words compared to unpredictable ones (\fig\ref{fig:supp_MEG_fig2_per_subj} ). These findings align with those of \cite{goldstein2022}, who used ECoG recordings to suggest that pre-onset encoding reflects predictions of the upcoming word. Moreover, as shown in \fig\ref{fig:FIR_filter}, the amplitude of the FIR filter used in the preprocessing step reaches zero within 100 ms, indicating that the observed pre-onset encoding is unlikely to result from temporal smearing introduced by the filter.

A potential concern is that pre-onset encoding might partly reflect residual processing of preceding words, since embeddings of neighboring words are highly correlated (see Fig. \ref{fig:supp_embedding_correlations}). 
To address this, we decorrelated the embeddings of nearby words using an approach similar to those used by \citet{goldstein2022} and \citet{schonmann2025stimulus}. However, rather than removing the effect of just one preceding word, we removed the influence of the eight previous words in the narrative (corresponding to $\sim$\,2 seconds before word onset). This procedure eliminates potential contributions from past-word encoding to the pre-onset encoding. As shown in \fig\ref{fig:fig2}.b, pre-onset encoding remains largely unchanged, suggesting that correlations between neighboring word embeddings cannot account for the observed pre-onset encoding. Notably, this decorrelation of embeddings also eliminates the baseline offset observed in the ECoG encoding, suggesting that this offset reflects encoding of shared representations among adjacent embeddings. 

To further investigate whether this pre-onset encoding genuinely reflects predictive pre-activation, we considered the possibility that the part of the pre-onset encoding might result from the encoding model learning word co-occurrences, particularly repeated bi-grams in the story. To test this, we re-ran the analysis after removing all but the first instance of each bi-gram. As shown in \fig\ref{fig:fig2}.b, this manipulation led to a slight reduction in pre-onset encoding, but the effect persisted. This finding suggests that pre-onset encoding is not merely an artifact of repeated co-occurrences in the stimulus set.

\subsection{Pre- and post-onset encoding do not reflect the same neural representations}

To assess whether pre-onset encoding truly reflects pre-activation of the same neural representations engaged after word onset, we compared the representations captured by the encoding model before and after word onset in more detail. Because separate linear regression models are fitted at each time point within the [-2, 2] s window in the encoding analyses above, it is not guaranteed that the model learns the same mapping between embeddings and brain activity across time. To directly test whether pre-onset encoding reflects predictive pre-activation, we evaluated whether the mapping between embeddings and the neural signal learned after word onset generalize to pre-onset activity using temporal generalization analysis \citep{king2014characterizing}. In this analysis, an encoder is trained on the neural signal at one time point and tested at another, relative to word onset. 

\begin{figure}
    \hspace{-1cm} 
    \centering
    \includegraphics[width=0.8\textwidth]{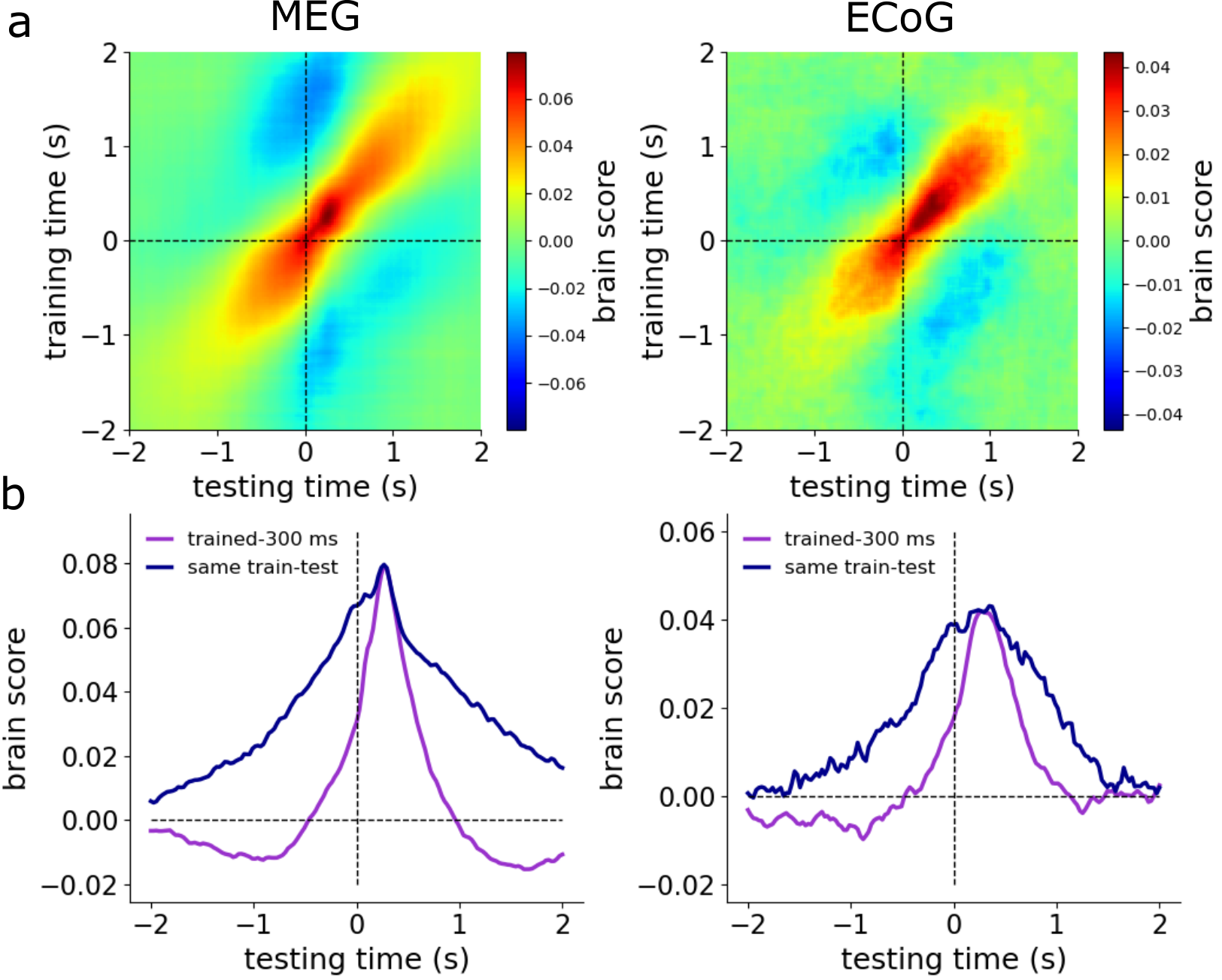}
    \caption{\textbf{Temporal generalization of representations captured by the encoding model differs before and after word onset.}
    \textbf{a.} Temporal generalization (TG) matrix computed by training the encoding model with decorrelated embeddings at one time point and testing it at another. Positive values indicate successful generalization of representations across time. The diagonal pattern reflects temporally dynamic rather than stable representations.
    \textbf{b.} Generalization profiles for models trained at the peak encoding response ($\sim$\,300 ms; purple) and along the diagonal (blue). The divergence between the two curves indicates that pre-onset encoding does not reflect the same representations engaged during word processing.}
    \label{fig:fig_TG}
\end{figure}

As shown in Fig. \ref{fig:fig_TG}.a, post-onset representations evolved dynamically over time, consistent with findings from \citet{gwilliams2022neural, gwilliams2025hierarchical}. TG matrices revealed limited generalization between pre- and post-onset periods: models trained at the peak of word representation ($\sim$\,300 ms) predicted pre-onset activity only within a narrow window immediately preceding word onset (\fig\ref{fig:fig_TG}.b). This overlap accounts for at most $\sim$\,400 ms of the pre-onset encoding observed previously, indicating that most of the pre-onset signal cannot be explained by pre-activation of the upcoming word.

\begin{figure}
    \hspace{-1cm} 
    \centering
    \includegraphics[width=0.8\textwidth]{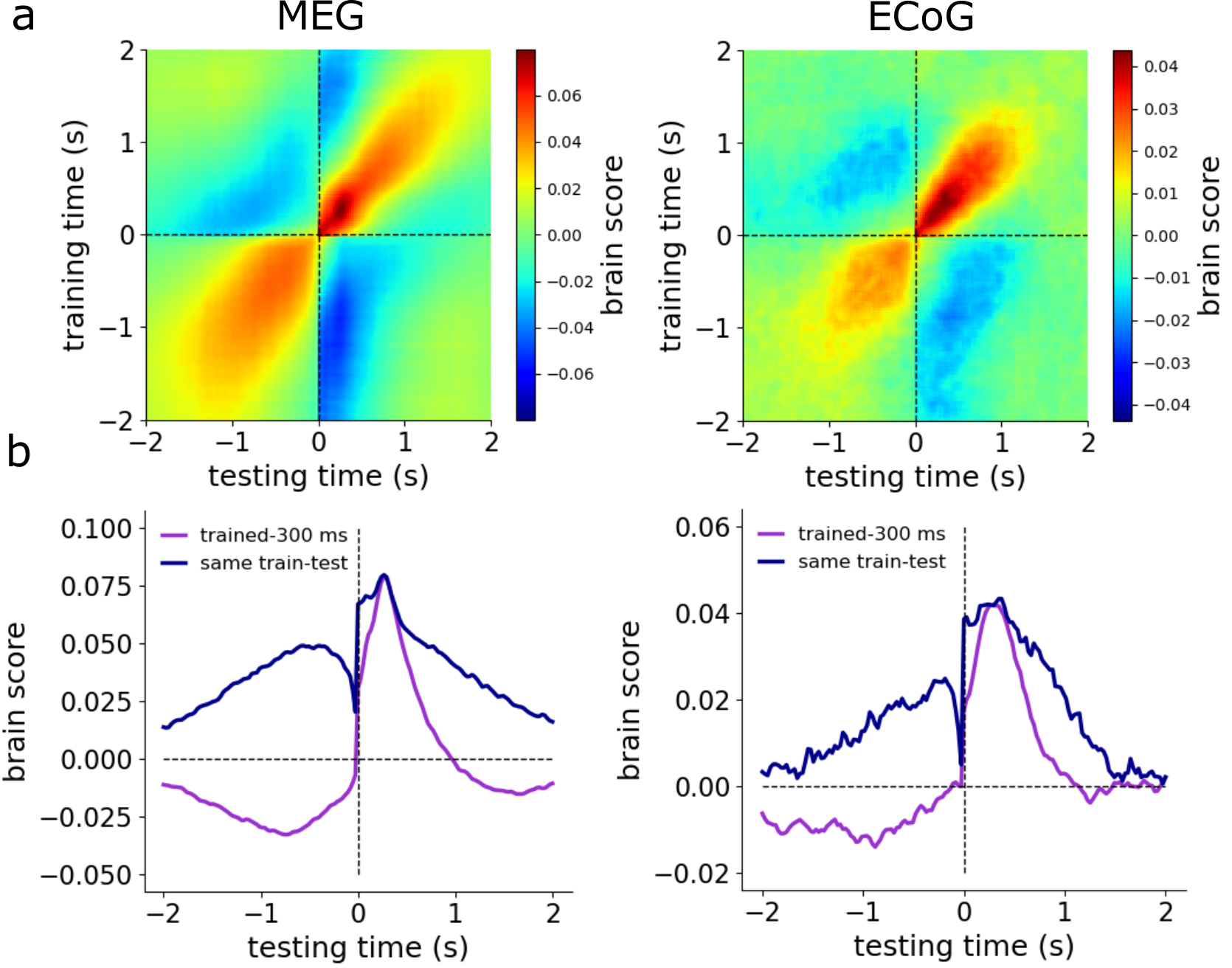}
    \caption{\textbf{Removing autocorrelation between pre- and post-onset activity does not eliminate pre-onset encoding.}
    \textbf{a.} Temporal generalization (TG) matrix computed using decorrelated embeddings after regressing out post-onset brain activity from the pre-onset signal. Removing correlations in the neural signal should, in principle, eliminate any trace of predictive pre-activation. Following this procedure, the small pre-onset generalization observed near word onset in \fig\ref{fig:fig_TG} disappears.
    \textbf{b.} Generalization profiles for models trained at the peak encoding response ($\sim$\,300 ms; purple) and along the diagonal (blue). The persistence of pre-onset encoding (blue curve) despite the absence of pre-activation indicates that pre-onset encoding is not necessarily a signature of prediction.}
    \label{fig:fig_TG_control}
\end{figure}

As an additional test of whether pre-onset encoding reflects pre-activation, we decorrelated pre-onset from post-onset activity (see \fig\ref{fig:supp_temporal_correlation} for the correlation profile). This step should remove any residual trace of genuine pre-activation of the features typically evoked by the word. After decorrelation (\fig\ref{fig:fig_TG_control}), pre-onset encoding remained largely intact, while the small temporal generalization observed before onset disappeared. 
Together, these results indicate that pre-onset encoding and post-onset processing rely on distinct neural representations, ruling out pre-activation as a shared source.

Interestingly, the temporal generalization (TG) matrix revealed negative generalization areas approximately one second from the diagonal (\fig\ref{fig:fig_TG}, \fig\ref{fig:fig_TG_control}), suggesting an inversion of the neural code over time. Similar negative–positive patterns have been observed in MEG studies of both visual and auditory processing \citep{carlson2013representational, king2014two, king2014characterizing} and have been interpreted as reflecting reversals in neural activity patterns. However, in our data, the negative values do not simply correspond to a sign reversal in the neural signal (see Supplementary Material, \fig\ref{fig:supp_correlation_word_subset}, for details).

\subsection{Future word embeddings do not improve encoding, while past word embeddings do}

\citet{caucheteux2023evidence} demonstrated that incorporating future word embeddings alongside the current word in the encoding model improves the model's performance in explaining fMRI BOLD signals, consistent with the idea that the brain pre-activates representations of upcoming words. They further observed a cortical hierarchy of predictive timescales, where lower-level areas have shorter prediction windows, while frontal areas anticipate further into the future.

Here we applied a similar approach as \citet{caucheteux2023evidence} to test whether including future word embeddings enhances model performance. Moreover, given the superior temporal resolution of these modalities, we can track this effect over time. We used the decorrelated embeddings to make sure there are no confounding effects from the nearby word embeddings sharing information (although the same pattern of results was observed with the original embeddings \fig\ref{fig:forecast_original}).

\begin{figure}
    \centering
    \includegraphics[width=1\textwidth]{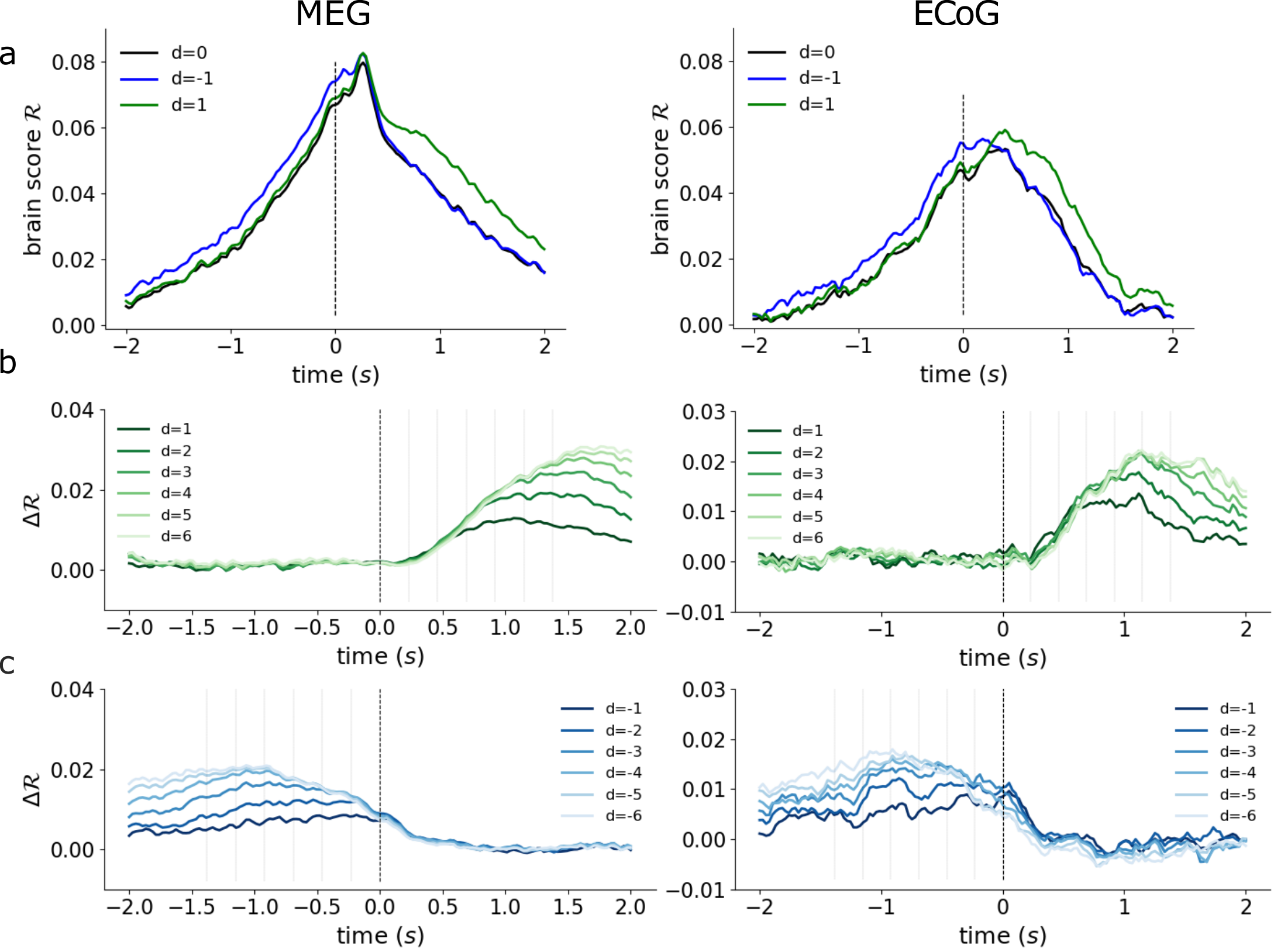}
    \caption{\textbf{Encoding of the future and past words.} 
    All curves represent averages across participants.
    The embedding vector is constructed by concatenating $d$ future word embeddings ($d > 0$) or $|d|$ past word embeddings ($d < 0$) along with the embedding of the current word $w_i$.
    \textbf{a} Including the next word embedding in the encoding model ($d=1$) enhances encoding only after that word is heard in the story, while including the previous word ($d=-1$) improves encoding even after the current word’s onset.
     Encoding enhancement, $\Delta\mathcal{R}$, is shown for \textbf{b} positive and \textbf{c} negative values of $d$. Vertical gray lines mark the median inter-word interval values. Adding each successive future word embedding improves encoding only after that word is heard in the narrative, while including previous words consistently improves encoding beyond their offset.
    }
    \label{fig:forecast}
\end{figure}

If the findings of \citet{caucheteux2023evidence} generalize to MEG and ECoG data, we would expect to observe an increased brain score ($\Delta R >0$) for models that include future word embeddings ($d>0$) around the onset of the current word. However, as shown in \fig\ref{fig:forecast}.a, the inclusion of the next word embedding ($d=1$) did not improve the brain score at the current word onset; rather, the increase appeared only about $250$ ms after the current word onset. Given that the average inter-word interval is 230-240 ms, this increase likely reflects the neural response to the next word once it has been heard.

Consistent with this, \fig\ref{fig:forecast}.c shows that increases in brain score emerge after approximately one inter-word interval, and curves for larger positive values of $d$ diverge at subsequent multiples of this interval. Together, these results indicate that future-word embeddings improve model performance only after the corresponding words are presented, arguing against pre-activation of upcoming word representations.

In contrast, including previous word embeddings enhances encoding performance at the current word onset (\fig\ref{fig:forecast}.b). This increase is most pronounced for the immediate previous word and diminishes with the inclusion of words further back in the past. This suggests that representations of previous words remain active for a period of time after they have been processed. \fig\ref{fig:forecast}.b and c clearly show an asymmetry between the representations of past and future words, with past words still being encoded in the MEG brain responses around the current word onset, while future words are not. 

To further examine whether this absence of effect could be attributed to our source selection method, we repeated the analysis using only MEG sources that, on average, exhibited enhanced encoding across the [0, 230] ms window for $d > 0$. Even with this strongly biased selection, encoding does not appear to improve noticeably within the [0, 230] ms interval (\fig\ref{fig:forecast_different_regions}). However, including previous word embeddings led to a measurable increase in encoding performance. The same is true when we specifically focus on the IFG electrodes in the ECoG data \fig\ref{fig:supp_IFG}, which have been previously found to robustly show pre-onset encoding \citep{goldstein2022}. Together these findings further support the idea that past words remain active while future words are not pre-encoded in MEG signals.

\section{Discussion}

Recent studies have leveraged the representational richness of LLMs to investigate predictive processing during natural speech comprehension \citep{goldstein2022, caucheteux2023evidence}. Here, we sought to replicate these findings across MEG and ECoG datasets and determine to what extent they may reflect predictive pre-activations of the upcoming words. 
 
Consistent with previous ECoG results \citep{goldstein2022}, we found that embeddings of upcoming words explained neural activity in language-related regions. Importantly, we ruled out several potential confounds. The pre-onset effects could not be attributed to residual encoding of the preceding word via similarity between adjacent embeddings, as suggested by \citet{schonmann2025stimulus}: orthogonalizing consecutive embeddings within the narrative did not diminish the pre-onset encoding. Moreover, similarly to \citet{goldstein2022}, removing all bigrams from the stimulus did not abolish the effect. Of course, removing n-grams does not remove all regularities inherent to natural speech, and fully removing all such dependencies (e.g., words that co-occur at longer distances) remains challenging as these issues are intrinsic to studying language in naturalistic contexts.

Given the many confounding factors that cannot be fully controlled, we directly examined the shared representations captured by the model before and after word onset. The TG analysis suggests that the observed pre-onset encoding does not primarily reflect predictive pre-activation of upcoming words. Although a small portion of the pre-onset signal shares representational similarity with post-onset activity, this overlap disappears once pre- and post-onset correlations are removed, while the overall pre-onset encoding remains. Thus, pre-onset encoding reflects a distinct mapping between embedding of the word $W_i$ and neural activity before its onset, rather than shared predictive representations.

Notably, the strongly diagonal structure of the TG matrix indicates that lexical representations are highly dynamic, with the mapping between embeddings and brain activity continuously transforming over time. This pattern is consistent with previous MEG findings during story listening \citep{gwilliams2022neural, gwilliams2025hierarchical} and has been proposed to reflect a neural coding strategy that allows the brain to process multiple stimuli simultaneously while avoiding interference between representations. This temporal variability underscores the need for a more nuanced view of what constitutes predictive pre-activation, as no stable representational state appears to persist after word presentation that could serve as its target.

To further assess whether previously reported effects truly reflect predictive pre-activation, we tested whether incorporating future word embeddings would enhance the encoding of neural responses to the current word ($w_i$). Unlike the fMRI findings of \citet{caucheteux2023evidence}, adding embeddings of upcoming words,  even the immediate next word ($w_{i+1}$), did not improve encoding performance. This result held after controlling for embedding correlations and when using static GloVe embeddings, which capture semantic content but lack contextual information. These findings are consistent with our earlier results from the TG analysis and argue against a predictive pre-activation account in which specific word features are the target of prediction.

Why, then, can encoding of future words be captured with fMRI but not with MEG or ECoG? fMRI measures slower hemodynamic responses associated with neuronal metabolic activity \citep{logothetis2008we}, integrating neural signals over relatively long temporal windows. As a result, it likely reflects integrative semantic processes operating across extended contexts. Moreover, to match fMRI’s temporal resolution, \citet{caucheteux2023evidence} averaged word embeddings within each repetition time (TR; approximately 1.5 s, covering about eight words). This temporal averaging likely produced embeddings that capture broader contextual meaning, resembling more stable, sentence-level representations rather than transient, word-specific ones. Consequently, the predictive signals observed in fMRI may primarily reflect pre-activation of contextual or semantic representations, rather than the prediction of specific upcoming words, a possibility that requires further investigation.

Unlike fMRI, MEG and ECoG rely on synchronized neural activity and with their high temporal resolution, directly capture rapid neural dynamics with millisecond precision. These modalities therefore primarily reflect instantaneous lexical processing rather than slower, integrative semantic processes. This interpretation aligns with findings from \citet{toneva2022combining}, who reported robust encoding of supra-word meaning (information derived from combinations of words beyond individual lexical items) in fMRI signals, whereas MEG predominantly captured lexical-level information. 

In our analysis inspired by \citet{caucheteux2023evidence}, we also tested whether incorporating embeddings of preceding words enhances MEG and ECoG encoding performance. Including past word embeddings  improved model performance, indicating that representations of prior words remain active over time (\fig\ref{fig:forecast}.b). This effect was robust across both GPT embeddings (before and after decorrelation) and static GloVe embeddings (\fig\ref{fig:forecast_glove}). These findings are consistent with those of \citet{toneva2022combining}, who reported that MEG signals encode both current and prior word meanings. Our replication of this pattern in ECoG further supports the view that electrophysiological recordings predominantly reflect representations of individual words rather than extended contextual predictions.

More broadly, our results support theories of postdictive processing, which propose that the brain buffers evidence over time, refining interpretations of earlier stimuli based on incoming information \citep{gwilliams2018spoken, hogendoorn2022perception,  szewczyk2022power}. These findings emphasize the need to clarify how electrophysiological measures encode both individual word representations and context-dependent information.

In summary, our findings challenge the interpretation that pre-onset encoding in electrophysiological signals reflects predictive pre-activation of upcoming words. Instead, they suggest that MEG and ECoG primarily capture transient lexical dynamics and the reverberation of prior information, consistent with postdictive processing. More broadly, our results call for a re-evaluation of what current encoding models actually capture and for a sharper definition of what counts as evidence for “prediction” in the brain.

\section{Funding}
Linda Geerligs was supported by a VIDI grant of the Netherlands Organization for Scientific Research (grant number VI.Vidi.201.150).

\section{Acknowledgements}
We would like to thank Lucia Rust for contributing to initial explorations of certain analyses; Nasir Ahmad and Dora Gözükara for the valuable and insightful discussions that greatly enhanced this work.

\section{Methods and Materials}

\subsection{MEG}

\subsubsection{Dataset}
We used an openly available dataset from \citet{armeni202210}. This dataset contains MEG recordings from a 275-channel axial gradiometer CTF system (sampling rate: $1200$ Hz) collected while three native English-speaking participants (1 female; aged 35, 30, and 28 years) passively listened to 10 stories from the Adventures of Sherlock Holmes. All participants were right-handed with no reported neurological, developmental or language deficits. This dataset provides long MEG recordings ($\sim$\,10 sessions of $\sim$\,1 hour each) per participant, enabling single-subject analysis. Additionally, participants wore MEG-compatible head casts to immobilize the head position during recording, yielding a high signal-to-noise ratio.  

\subsubsection{Preprocessing}

All data were preprocessed using the MNE-Python open-source software package, v. 1.6.1 \citep{gramfort2013meg,larson_2024_10519948}. Because our analysis approach required continuous data, we did not remove any noisy segments from the data; instead, we pre-processed the data using ICA, separately for each session and participant, removing components that related to eye movement and heartbeat artifacts. The components were identified from raw data that was filtered between 0.5 and 40~Hz using an FIR filter and then resampled to 150 Hz. After identifying the components, we removed them from the data filtered between 0.1-40 Hz using a zero-phase non-causal FIR filter. 

\subsubsection{Source reconstruction}

Using the FreeSurfer software ~\citep{dale1999cortical}, we computed a single layer boundary element model for each participant using the 3T MRI anatomical scans provided in the dataset. For each subject, an individual single-shell boundary element model (BEM) was made and individual forward models were calculated. Source reconstruction was performed using a linearly constrained minimum variance (LCMV) beamformer with unit-noise-gain, optimizing source orientation to maximize output power \citep{sekihara2008adaptive,VanVeen,westner2022unified}.

The LCMV adaptive filter was computed using a data covariance matrix within the [0, 150] ms window after the stimulus onset (each word in the story). We regularized the covariance using Tikhonov regularization with 10\% of overall sensor power. Finally, the filter was applied to the broadband raw time series. We computed 150 source regions based on the Destrieux Atlas (available through FreeSurfer; \citep{dale1999cortical,destrieux2010automatic,fischl2004automatically}), averaging source points within regions for each participant.

\subsection{ECoG dataset and preprocessing}
Here we use on the “Podcast” ECoG dataset introduced by \citet{zada2025podcast}. The dataset contains ECoG recordings from 9 epilepsy patients with a total of 1,330 electrodes (grid, strip, and depth) as they listened to a ~30-min of "This American Life" podcast. The open dataset provides preprocessed data with bad-channel exclusion, artifact removal, and high-gamma (70–200 Hz) broadband power extraction. We used the provided high-gamma signal directly, applying no additional preprocessing except for downsampling from 512 Hz to 150 Hz to match the MEG sampling frequency.

\subsection{Word embeddings}

To extract contextual word embeddings, we used the pre-trained GPT2-small model with 12 hidden layers parameters \citep{radford2019language}. We first tokenized the text (without punctuation and capitalization) using the model tokenizer. To obtain the GPT embedding for each token $w_i$ in the story, we fed chunks of the tokenized story in windows of size 50 tokens $[w_{i-49},w_{i-48},...,w_{i}]$ (with the first 50 words having a smaller context window) to the GPT2-small tokenizer model and extracted the hidden state of the 8th out of 12 layers of the model $\mathbf{x} \in \mathbb{R}^{768}$ corresponding to the last word. This was because the 8th layer showed slightly higher encoding than the 12th layer. Finally, words with multiple tokens assigned by the model were excluded to remove within-word effects. This left us with a total of 80472 words split into 10 similar-sized sessions for the MEG  dataset and 4708 in a single session for the ECoG dataset. For certain conditions, bi-grams were removed from the story which reduced the MEG dataset to ~2,5 h and the ECoG dataset ~19 min. 
For non-contextual embeddings, we used GloVe (Wikipedia 2014) embeddings \citep{pennington2014glove} with vectors of size $300$. To reduce computational load, the embeddings were dimensionally reduced to $50$ using PCA. This was done on the training set and applied to the test set to prevent data leakage. This resulted in  $\mathbf{X} \in \mathbb{R}^{N \times 50} $ as predictors.

\subsection{Encoding model and brain score}

To preprocess the MEG data for encoding model training, time series data were smoothed using a 100 ms rectangular window \citep[as in][]{goldstein2022}. The data were then epoched into time windows from [-2 to +2 sec] around word onset and downsampled to 151 time points within this range, yielding a target matrix $\mathbf{Y} \in \mathbb{R}^{N \times 151}$, where $N$ is the number of words.

The encoding model was built using linear ridge-regression with word embeddings $\mathbf{X}$ as predictors and brain response $\mathbf{Y}$ as targets. A separate model was fit for each time point $\mathbf{t}$ within the [-2 to +2 sec] window. To find the optimal regularization parameter $\mathbf{\lambda}$ of the ridge-regression, we used grid search with generalized cross-validation, which approximates leave-one-out cross-validation \citep{hastie2017elements}. We defined a grid to search by first sampling the effective degrees of freedom of the ridge-regression fit from $[1, N]$, since its parameter space is bounded from above. To find degrees of freedom of the ridge regression we used singular value decomposition $\mathbf{X}=\mathbf{U}\mathbf{S}\mathbf{V}^T$ where $\mathbf{U}$ and $\mathbf{V}$ are orthogonal and $\mathbf{S}$ diagonal matrices. The degrees of freedom is then defined as $df(\lambda_i) = \sum_{j=1}^N  \frac{s_j^2}{s_j^2+\lambda_i}$.
We then used Newton’s method to solve $df$ for $\lambda_i$. Once the grid was defined, we chose the optimal value $\lambda^\ast$ that minimizes the generalized cross-validation error using Leave-One-Out Cross-Validation implementation of scikit-learn library in python \citep{scikit_learn}. As $\lambda^\ast$ remained highly stable across timepoints, electrode/sources, sessions and participants, a fixed value was used per dataset in subsequent computations to improve efficiency.

We assessed model performance using 5-fold cross-validation, training on 80\% of the story and testing on the remaining 20\%  Training data was normalized to a mean of 0 and variance of 1, and the same transformation applied to the test set. To prevent overfitting due to temporal autocorrelation in brain activity, training and test sets were temporally separated \citep{feghhi2024large}. This produced a predicted brain signal for each word at each time point within the [-2 to +2 sec] window around word onset.
To evaluate model fit, we computed the Pearson correlation between the predicted and actual brain signals at each time point $t$ around word onset across all words to compute the brain score $\mathcal{R}$. 

\subsection{Temporal generalization analysis}
To examine how similar representations captured by the encoding model are pre- and post-onset, similarly to before, we trained an encoder to predict brain responses at time point $t_i$. However this time, we tested the model by predicting brain responses at time point $t_{i+n}$ and calculated Pearson correlation between this prediction and actual brain response (i.e. brain score).  Doing so for all 151 time points yielded a temporal generalization matrix of size $151\times 151$. 

\subsection{Analysis with expanded future/past window}

To test whether future/past words are represented in the brain response, we constructed embedding vectors by concatenating additional word embeddings to the current word embedding. For a given window size $d$, we defined the embedding as $\mathbf{z}^d_i=\mathbf{x}_{i+d}\oplus\mathbf{x}_{i+d-1}\oplus ... \oplus \mathbf{x}_{i}$ where $\mathbf{x}_i$ is the word embedding vector of word $w_i$. Negative and positive $d$ values correspond to windows spanning the words in past or future respectively.  

Finally, we reduced dimensions of both $\mathbf{z}^d_i$ and $\mathbf{x}_i$ to $50$ and concatenated the two as $\hat{\mathbf{z}}^d_i \oplus \hat{\mathbf{x}}_i$, where $\hat{\mathbf{z}}$ and $\hat{\mathbf{x}}$ are the vectors after applying PCA. 
To see how much encoding improves, we computed the change in the brain score when including future/past word embeddings as  $\Delta\mathcal{R} = \mathcal{R}(\hat{\mathbf{Z}}^d \oplus \hat{\mathbf{X}})- \mathcal{R}(\hat{\mathbf{X}})$, where $\mathcal{R(.)}$ denotes the brain score computed given the respective embeddings.

\subsection{MEG Source selection}

To compute the brain-wide brain score, we selected a subset of 30 sources for each participant. To avoid double dipping, we ensured that the source selection for each participant was not based on their own data. To do this, we averaged the brain score of two participants for each MEG source and chose the top 30 sources with the highest values for the third participant.

\section{Data and code availability}
The Python code used in this paper can be found at \url{https://github.com/sahelazizpour/Linguistic_Preactivation}

\bibliography{elife-sample}

@article{armeni202210,
  title={A 10-hour within-participant magnetoencephalography narrative dataset to test models of language comprehension},
  author={Armeni, Kristijan and G{\"u}{\c{c}}l{\"u}, Umut and van Gerven, Marcel and Schoffelen, Jan-Mathijs},
  journal={Scientific Data},
  volume={9},
  number={1},
  pages={278},
  year={2022},
  publisher={Nature Publishing Group UK London}
}

@article{goldstein2022,
  title={Shared computational principles for language processing in humans and deep language models},
  author={Goldstein, Ariel and Zada, Zaid and Buchnik, Eliav and Schain, Mariano and Price, Amy and Aubrey, Bobbi and Nastase, Samuel A and Feder, Amir and Emanuel, Dotan and Cohen, Alon and others},
  journal={Nature neuroscience},
  volume={25},
  number={3},
  pages={369--380},
  year={2022},
  publisher={Nature Publishing Group US New York}
}

@article{friston2010free,
  title={The free-energy principle: a unified brain theory?},
  author={Friston, Karl},
  journal={Nature reviews neuroscience},
  volume={11},
  number={2},
  pages={127--138},
  year={2010},
  publisher={Nature publishing group}
}

@article{clark2013whatever,
  title={Whatever next? Predictive brains, situated agents, and the future of cognitive science},
  author={Clark, Andy},
  journal={Behavioral and brain sciences},
  volume={36},
  number={3},
  pages={181--204},
  year={2013},
  publisher={Cambridge University Press}
}

@article{willems2020narratives,
  title={Narratives for neuroscience},
  author={Willems, Roel M and Nastase, Samuel A and Milivojevic, Branka},
  journal={Trends in neurosciences},
  volume={43},
  number={5},
  pages={271--273},
  year={2020},
  publisher={Elsevier}
}

@article{de2018expectations,
  title={How do expectations shape perception?},
  author={De Lange, Floris P and Heilbron, Micha and Kok, Peter},
  journal={Trends in cognitive sciences},
  volume={22},
  number={9},
  pages={764--779},
  year={2018},
  publisher={Elsevier}
}

@article{schrimpf2021neural,
  title={The neural architecture of language: Integrative modeling converges on predictive processing},
  author={Schrimpf, Martin and Blank, Idan Asher and Tuckute, Greta and Kauf, Carina and Hosseini, Eghbal A and Kanwisher, Nancy and Tenenbaum, Joshua B and Fedorenko, Evelina},
  journal={Proceedings of the National Academy of Sciences},
  volume={118},
  number={45},
  pages={e2105646118},
  year={2021},
  publisher={National Acad Sciences}
}

@article{caucheteux2022brains,
  title={Brains and algorithms partially converge in natural language processing},
  author={Caucheteux, Charlotte and King, Jean-R{\'e}mi},
  journal={Communications biology},
  volume={5},
  number={1},
  pages={134},
  year={2022},
  publisher={Nature Publishing Group UK London}
}

@article{zada2024shared,
  title={A shared model-based linguistic space for transmitting our thoughts from brain to brain in natural conversations},
  author={Zada, Zaid and Goldstein, Ariel and Michelmann, Sebastian and Simony, Erez and Price, Amy and Hasenfratz, Liat and Barham, Emily and Zadbood, Asieh and Doyle, Werner and Friedman, Daniel and others},
  journal={Neuron},
  volume={112},
  number={18},
  pages={3211--3222},
  year={2024},
  publisher={Elsevier}
}

@article{federmeier2007thinking,
  title={Thinking ahead: The role and roots of prediction in language comprehension},
  author={Federmeier, Kara D},
  journal={Psychophysiology},
  volume={44},
  number={4},
  pages={491--505},
  year={2007},
  publisher={Wiley Online Library}
}

@article{kutas2011look,
  title={A look around at what lies ahead: Prediction and predictability in language processing},
  author={Kutas, Marta and DeLong, Katherine A and Smith, Nathaniel J},
  journal={Predictions in the brain: Using our past to generate a future},
  volume={190207},
  number={10.1093},
  year={2011}
}

@article{traxler2000effects,
  title={Effects of sentence constraint on priming in natural language comprehension.},
  author={Traxler, Matthew J and Foss, Donald J},
  journal={Journal of Experimental Psychology: Learning, Memory, and Cognition},
  volume={26},
  number={5},
  pages={1266},
  year={2000},
  publisher={American Psychological Association}
}

@article{smith2013effect,
  title={The effect of word predictability on reading time is logarithmic},
  author={Smith, Nathaniel J and Levy, Roger},
  journal={Cognition},
  volume={128},
  number={3},
  pages={302--319},
  year={2013},
  publisher={Elsevier}
}

@article{magyari2012prediction,
  title={Prediction of turn-ends based on anticipation of upcoming words},
  author={Magyari, Lilla and De Ruiter, Jan P},
  journal={Frontiers in psychology},
  volume={3},
  pages={376},
  year={2012},
  publisher={Frontiers Media SA}
}

@article{de2006projecting,
  title={Projecting the end of a speaker's turn: A cognitive cornerstone of conversation},
  author={De Ruiter, Jan-Peter and Mitterer, Holger and Enfield, Nick J},
  journal={Language},
  volume={82},
  number={3},
  pages={515--535},
  year={2006},
  publisher={Linguistic Society of America}
}

@article{kutas2011thirty,
  title={Thirty years and counting: finding meaning in the N400 component of the event-related brain potential (ERP)},
  author={Kutas, Marta and Federmeier, Kara D},
  journal={Annual review of psychology},
  volume={62},
  number={1},
  pages={621--647},
  year={2011},
  publisher={Annual Reviews}
}

@article{delong2005probabilistic,
  title={Probabilistic word pre-activation during language comprehension inferred from electrical brain activity},
  author={DeLong, Katherine A and Urbach, Thomas P and Kutas, Marta},
  journal={Nature neuroscience},
  volume={8},
  number={8},
  pages={1117--1121},
  year={2005},
  publisher={Nature Publishing Group US New York}
}

@article{kuperberg2016we,
  title={What do we mean by prediction in language comprehension?},
  author={Kuperberg, Gina R and Jaeger, T Florian},
  journal={Language, cognition and neuroscience},
  volume={31},
  number={1},
  pages={32--59},
  year={2016},
  publisher={Taylor \& Francis}
}

@article{hamilton2020revolution,
  title={The revolution will not be controlled: natural stimuli in speech neuroscience},
  author={Hamilton, Liberty S and Huth, Alexander G},
  journal={Language, cognition and neuroscience},
  volume={35},
  number={5},
  pages={573--582},
  year={2020},
  publisher={Taylor \& Francis}
}

@article{toneva2022combining,
  title={Combining computational controls with natural text reveals aspects of meaning composition},
  author={Toneva, Mariya and Mitchell, Tom M and Wehbe, Leila},
  journal={Nature computational science},
  volume={2},
  number={11},
  pages={745--757},
  year={2022},
  publisher={Nature Publishing Group US New York}
}

@article{caucheteux2023evidence,
  title={Evidence of a predictive coding hierarchy in the human brain listening to speech},
  author={Caucheteux, Charlotte and Gramfort, Alexandre and King, Jean-R{\'e}mi},
  journal={Nature human behaviour},
  volume={7},
  number={3},
  pages={430--441},
  year={2023},
  publisher={Nature Publishing Group UK London}
}

@article{gwilliams2022neural,
  title={Neural dynamics of phoneme sequences reveal position-invariant code for content and order},
  author={Gwilliams, Laura and King, Jean-Remi and Marantz, Alec and Poeppel, David},
  journal={Nature communications},
  volume={13},
  number={1},
  pages={6606},
  year={2022},
  publisher={Nature Publishing Group UK London}
}

@article{gramfort2013meg,
  title={MEG and EEG data analysis with MNE-Python},
  author={Gramfort, Alexandre and Luessi, Martin and Larson, Eric and Engemann, Denis A and Strohmeier, Daniel and Brodbeck, Christian and Goj, Roman and Jas, Mainak and Brooks, Teon and Parkkonen, Lauri and others},
  journal={Frontiers in Neuroinformatics},
  volume={7},
  pages={267},
  year={2013},
  publisher={Frontiers Media SA}
}

@article{dale1999cortical,
  title={Cortical surface-based analysis: I. Segmentation and surface reconstruction},
  author={Dale, Anders M and Fischl, Bruce and Sereno, Martin I},
  journal={Neuroimage},
  volume={9},
  number={2},
  pages={179--194},
  year={1999},
  publisher={Elsevier}
}

@ARTICLE{VanVeen,
  author={Van Veen, B.D. and Van Drongelen, W. and Yuchtman, M. and Suzuki, A.},
  journal={IEEE Transactions on Biomedical Engineering}, 
  title={Localization of brain electrical activity via linearly constrained minimum variance spatial filtering}, 
  year={1997},
  volume={44},
  number={9},
  pages={867-880},
  doi={10.1109/10.623056}}

@article{radford2019language,
  title={Language models are unsupervised multitask learners},
  author={Radford, Alec and Wu, Jeffrey and Child, Rewon and Luan, David and Amodei, Dario and Sutskever, Ilya and others},
  journal={OpenAI blog},
  volume={1},
  number={8},
  pages={9},
  year={2019}
}

@misc{hastie2017elements,
  title={The elements of statistical learning: data mining, inference, and prediction},
  author={Hastie, Trevor and Tibshirani, Robert and Friedman, Jerome},
  year={2017},
  publisher={Springer}
}

@article{feghhi2024large,
  title={What Are Large Language Models Mapping to in the Brain? A Case Against Over-Reliance on Brain Scores},
  author={Feghhi, Ebrahim and Hadidi, Nima and Song, Bryan and Blank, Idan A and Kao, Jonathan C},
  journal={arXiv preprint arXiv:2406.01538},
  year={2024}
}

@inproceedings{pennington2014glove,
  title={Glove: Global vectors for word representation},
  author={Pennington, Jeffrey and Socher, Richard and Manning, Christopher D},
  booktitle={Proceedings of the 2014 conference on empirical methods in natural language processing (EMNLP)},
  pages={1532--1543},
  year={2014}
}

@article{antonello2024predictive,
  title={Predictive coding or just feature discovery? An alternative account of why language models fit brain data},
  author={Antonello, Richard and Huth, Alexander},
  journal={Neurobiology of Language},
  volume={5},
  number={1},
  pages={64--79},
  year={2024},
  publisher={MIT Press One Broadway, 12th Floor, Cambridge, Massachusetts 02142, USA~…}
}

@article{hogendoorn2022perception,
  title={Perception in real-time: predicting the present, reconstructing the past},
  author={Hogendoorn, Hinze},
  journal={Trends in Cognitive Sciences},
  volume={26},
  number={2},
  pages={128--141},
  year={2022},
  publisher={Elsevier}
}

@article{gwilliams2018spoken,
  title={In spoken word recognition, the future predicts the past},
  author={Gwilliams, Laura and Linzen, Tal and Poeppel, David and Marantz, Alec},
  journal={Journal of Neuroscience},
  volume={38},
  number={35},
  pages={7585--7599},
  year={2018},
  publisher={Soc Neuroscience}
}

@article{goldstein2022correspondence,
  title={Correspondence between the layered structure of deep language models and temporal structure of natural language processing in the human brain},
  author={Goldstein, Ariel and Ham, Eric and Nastase, Samuel A and Zada, Zaid and Grinstein-Dabus, Avigail and Aubrey, Bobbi and Schain, Mariano and Gazula, Harshvardhan and Feder, Amir and Doyle, Werner and others},
  journal={BioRxiv},
  pages={2022--07},
  year={2022},
  publisher={Cold Spring Harbor Laboratory}
}

@article{szewczyk2022power,
  title={The power of “good”: Can adjectives rapidly decrease as well as increase the availability of the upcoming noun?},
  author={Szewczyk, Jakub M and Mech, Emily N and Federmeier, Kara D},
  journal={Journal of Experimental Psychology: Learning, Memory, and Cognition},
  volume={48},
  number={6},
  pages={856},
  year={2022},
  publisher={American Psychological Association}
}

@incollection{baroni2022proper,
  title={On the proper role of linguistically oriented deep net analysis in linguistic theorising},
  author={Baroni, Marco},
  booktitle={Algebraic structures in natural language},

  year={2022},
  publisher={CRC Press}
}

@article{wang2024dissociating,
  title={Dissociating the pre-activation of word meaning and form during sentence comprehension: Evidence from EEG representational similarity analysis},
  author={Wang, Lin and Brothers, Trevor and Jensen, Ole and Kuperberg, Gina R},
  journal={Psychonomic Bulletin \& Review},
  volume={31},
  number={2},
  pages={862--873},
  year={2024},
  publisher={Springer}
}

@article{wang2020neural,
  title={Neural evidence for the prediction of animacy features during language comprehension: evidence from MEG and EEG representational similarity analysis},
  author={Wang, Lin and Wlotko, Edward and Alexander, Edward and Schoot, Lotte and Kim, Minjae and Warnke, Lena and Kuperberg, Gina R},
  journal={Journal of Neuroscience},
  volume={40},
  number={16},
  pages={3278--3291},
  year={2020},
  publisher={Soc Neuroscience}
}

@article{destrieux2010automatic,
  title={Automatic parcellation of human cortical gyri and sulci using standard anatomical nomenclature},
  author={Destrieux, Christophe and Fischl, Bruce and Dale, Anders and Halgren, Eric},
  journal={Neuroimage},
  volume={53},
  number={1},
  pages={1--15},
  year={2010},
  publisher={Elsevier}
}

@article{fischl2004automatically,
  title={Automatically parcellating the human cerebral cortex},
  author={Fischl, Bruce and Van Der Kouwe, Andr{\'e} and Destrieux, Christophe and Halgren, Eric and S{\'e}gonne, Florent and Salat, David H and Busa, Evelina and Seidman, Larry J and Goldstein, Jill and Kennedy, David and others},
  journal={Cerebral cortex},
  volume={14},
  number={1},
  pages={11--22},
  year={2004},
  publisher={Oxford University Press}
}

@book{sekihara2008adaptive,
  title={Adaptive spatial filters for electromagnetic brain imaging},
  author={Sekihara, Kensuke and Nagarajan, Srikatan S},
  year={2008},
  publisher={Springer Science \& Business Media}
}

@article{westner2022unified,
  title={A unified view on beamformers for M/EEG source reconstruction},
  author={Westner, Britta U and Dalal, Sarang S and Gramfort, Alexandre and Litvak, Vladimir and Mosher, John C and Oostenveld, Robert and Schoffelen, Jan-Mathijs},
  journal={NeuroImage},
  volume={246},
  pages={118789},
  year={2022},
  publisher={Elsevier}
}

@article{scikit_learn,
  title={Scikit-learn: Machine Learning in {P}ython},
  author={Pedregosa, F. and Varoquaux, G. and Gramfort, A. and Michel, V.
          and Thirion, B. and Grisel, O. and Blondel, M. and Prettenhofer, P.
          and Weiss, R. and Dubourg, V. and Vanderplas, J. and Passos, A. and
          Cournapeau, D. and Brucher, M. and Perrot, M. and Duchesnay, E.},
  journal={Journal of Machine Learning Research},
  volume={12},
  pages={2825--2830},
  year={2011}
}

@software{larson_2024_10519948,
  author       = {Larson, Eric and
                  Gramfort, Alexandre and
                  Engemann, Denis A and
                  Leppakangas, Jaakko and
                  Brodbeck, Christian and
                  Jas, Mainak and
                  Brooks, Teon and
                  Sassenhagen, Jona and
                  Luessi, Martin and
                  McCloy, Daniel and
                  King, Jean-Remi and
                  Höchenberger, Richard and
                  Goj, Roman and
                  Favelier, Guillaume and
                  Brunner, Clemens and
                  van Vliet, Marijn and
                  Wronkiewicz, Mark and
                  Holdgraf, Chris and
                  Rockhill, Alex and
                  Massich, Joan and
                  Bekhti, Yousra and
                  Scheltienne, Mathieu and
                  Appelhoff, Stefan and
                  Leggitt, Alan and
                  Dykstra, Andrew and
                  Luke, Rob and
                  Trachel, Romain and
                  De Santis, Lorenzo and
                  Panda, Asish and
                  Magnuski, Mikołaj and
                  Westner, Britta and
                  Billinger, Martin and
                  Wakeman, Dan G and
                  Strohmeier, Daniel and
                  Bharadwaj, Hari and
                  Linzen, Tal and
                  Barachant, Alexandre and
                  Ruzich, Emily and
                  Bailey, Christopher J and
                  Li, Adam and
                  Moutard, Clément and
                  Bloy, Luke and
                  Raimondo, Fede and
                  Nurminen, Jussi and
                  Montoya, Jair and
                  Woodman, Marmaduke and
                  Lee, Ingoo and
                  Schulz, Martin and
                  Foti, Nick and
                  Nangini, Cathy and
                  García Alanis, José C and
                  Huberty, Scott and
                  Hauk, Olaf and
                  Maddox, Ross and
                  Orfanos, Dimitri Papadopoulos and
                  LaPlante, Roan and
                  Drew, Ashley and
                  Dinh, Christoph and
                  Dumas, Guillaume and
                  Benerradi, Johann and
                  Hartmann, Thomas and
                  Ort, Eduard and
                  Pasler, Paul and
                  Repplinger, Stefan and
                  Rudiuk, Alexander and
                  Radanovic, Ana and
                  Buran, Brad and
                  Massias, Mathurin and
                  Hämäläinen, Matti and
                  Sripad, Praveen and
                  Chirkov, Valerii and
                  Mullins, Christopher and
                  Raimundo, Félix and
                  Alday, Phillip and
                  Pari, Ram and
                  Kornblith, Simon and
                  Halchenko, Yaroslav and
                  Luo, Yu-Han and
                  Kasper, Johannes and
                  Doelling, Keith and
                  Jensen, Mads and
                  Gahlot, Tanay and
                  Nunes, Adonay and
                  Gramfort, Alexandre and
                  Gütlin, Dirk and
                  kjs and
                  Weinstein, Alejandro and
                  Lamus, Camilo and
                  Galván, Catalina María and
                  Moënne-Loccoz, Cristóbal and
                  Altukhov, Dmitrii and
                  Peterson, Erica and
                  Heinila, Erkka and
                  Hanna, Jevri and
                  Houck, Jon and
                  Kaneda, Michiru and
                  Klein, Natalie and
                  Roujansky, Paul and
                  Kern, Simon and
                  Rantala, Antti and
                  Maess, Burkhard and
                  O'Reilly, Christian and
                  Kolkhorst, Henrich and
                  Banville, Hubert and
                  Zhang, Jack and
                  Woessner, Jacob and
                  Maksymenko, Kostiantyn and
                  Clarke, Maggie and
                  Anelli, Matteo and
                  Chapochnikov, Nikolai and
                  Bannier, Pierre-Antoine and
                  Choudhary, Saket and
                  Forster, Carina and
                  Kim, Cora and
                  Klotzsche, Felix and
                  Wong, Fu-Te and
                  Kojcic, Ivana and
                  Nielsen, Jesper Duemose and
                  Lankinen, Kaisu and
                  Tabavi, Kambiz and
                  Thibault, Louis and
                  Gerster, Moritz and
                  Gayraud, Nathalie and
                  Ward, Nick and
                  Ruuskanen, Santeri and
                  Radanovic, Ana and
                  Quinn, Andrew and
                  Gauthier, Antoine and
                  Pinsard, Basile and
                  Welke, Dominik and
                  Welke, Dominik and
                  Stephen, Emily and
                  Hornberger, Erik and
                  Hathaway, Evan and
                  Kalenkovich, Evgenii and
                  Mamashli, Fahimeh and
                  Marinato, Giorgio and
                  Anevar, Hafeza and
                  Sosulski, Jan and
                  Stout, Jeff and
                  Calder-Travis, Joshua and
                  Eisenman, Larry and
                  Esch, Lorenz and
                  Dovgialo, Marian and
                  Barascud, Nicolas and
                  Legrand, Nicolas and
                  Falach, Rotem and
                  Deslauriers-Gauthier, Samuel and
                  Cotroneo, Silvia and
                  Matindi, Steve and
                  Bierer, Steven and
                  Férat, Victor and
                  Peterson, Victoria and
                  Baratz, Zvi and
                  Tonin, Alessandro and
                  Kovrig, Alexander and
                  Pascarella, Annalisa and
                  Karekal, Apoorva and
                  de la Torre, Carlos and
                  Gohil, Chetan and
                  Zhao, Christina and
                  Krzemiński, Dominik and
                  Makowski, Dominique and
                  Mikulan, Ezequiel and
                  Belonosov, Gennadiy and
                  O'Neill, George and
                  Abdelhedi, Hamza and
                  Schiratti, Jean-Baptiste and
                  Evans, Jen and
                  Veillette, John and
                  Drew, Jordan and
                  Teves, Joshua and
                  Zhu, Judy D and
                  Armeni, Kristijan and
                  Mathewson, Kyle and
                  Gwilliams, Laura and
                  Varghese, Lenny and
                  Gemein, Lukas and
                  Hecker, Lukas and
                  Lx37 and
                  van Es, Mats and
                  Boggess, Matt and
                  Eberlein, Matthias and
                  Sherif, Mohamed and
                  Kozhemiako, Nataliia and
                  Srinivasan, Naveen and
                  Wilming, Niklas and
                  Kozynets, Oleh and
                  Ablin, Pierre and
                  Bertrand, Quentin and
                  Shoorangiz, Reza and
                  Hübner, Rodrigo and
                  Sommariva, Sara and
                  Er, Sena and
                  Khan, Sheraz and
                  Herbst, Sophie and
                  Datta, Sumalyo and
                  Papadopoulo, Theodore and
                  Jochmann, Thomas and
                  Binns, Thomas Samuel and
                  Merk, Timon and
                  Flak, Tod and
                  Dupré la Tour, Tom and
                  Stenner, Tristan and
                  NessAiver, Tziona and
                  akshay0724 and
                  sviter and
                  Earle-Richardson, Aaron and
                  Hindle, Abram and
                  Koutsou, Achilleas and
                  Fecker, Adeline and
                  Wagner, Adina and
                  Ciok, Alex and
                  Gilbert, Andy and
                  Pradhan, Aniket and
                  Padee, Anna and
                  Dubarry, Anne-Sophie and
                  Waniek, Anton Nikolas and
                  Singhal, Archit and
                  Rokem, Ariel and
                  Pelzer, Arne and
                  Hurst, Austin and
                  Beasley, Ben and
                  Nicenboim, Bruno and
                  de la Torre, Carlos and
                  Clauss, Christian and
                  Mista, Christian and
                  Li, Chun-Hui and
                  Braboszcz, Claire and
                  Schad, Daniel Carlström and
                  Hasegan, Daniel and
                  Tse, Daniel and
                  Sleiter, Darin Erat and
                  Haslacher, David and
                  Sabbagh, David and
                  Kostas, Demetres and
                  Petkova, Desislava and
                  Issagaliyeva, Dinara and
                  Das, Diptyajit and
                  Wetzel, Dominik and
                  Eich, Eberhard and
                  DuPre, Elizabeth and
                  Lau, Ellen and
                  Olivetti, Emanuele and
                  Varano, Enrico and
                  Altamiranda, Enzo and
                  Brayet, Eric and
                  de Montalivet, Etienne and
                  Goldstein, Evgeny and
                  Zamberlan, Federico and
                  Pop, Florin and
                  Weber, Frederik D and
                  Tan, Gansheng and
                  Brookshire, Geoff and
                  O'Neill, George and
                  Giulio and
                  Reina, Gonzalo and
                  Maymandi, Hamid and
                  Sonntag, Hermann and
                  Ye, Hongjiang and
                  Shin, Hyonyoung and
                  Elmas, Hüseyin Orkun and
                  Machairas, Ilias and
                  Skelin, Ivan and
                  Zubarev, Ivan and
                  Kaczmarzyk, Jakub and
                  Zerfowski, Jan and
                  van den Bosch, Jasper J F and
                  Van Der Donckt, Jeroen and
                  van der Meer, Johan and
                  Niediek, Johannes and
                  Koen, Josh and
                  Bear, Joshua J and
                  Dammers, Juergen and
                  Galán, Julia Guiomar Niso and
                  Welzel, Julius and
                  Slama, Katarina and
                  Leinweber, Katrin and
                  Grabot, Laetitia and
                  Andersen, Lau Møller and
                  Barbosa, Leonardo S and
                  Hamilton, Liberty and
                  Alfine, Lorenzo and
                  Hejtmánek, Lukáš and
                  Balatsko, Maksym and
                  Kitzbichler, Manfred and
                  Kumar, Manoj and
                  Kadwani, Manorama and
                  Sutela, Manu and
                  Koculak, Marcin and
                  Henney, Mark Alexander and
                  Schulz, Martin and
                  van Harmelen, Martin and
                  MartinBaBer and
                  Courtemanche, Matt and
                  Tucker, Matt and
                  Visconti di Oleggio Castello, Matteo and
                  Dold, Matthias and
                  Toivonen, Matti and
                  Shader, Maureen and
                  Cespedes, Mauricio and
                  Krause, Michael and
                  Rybář, Milan and
                  He, Mingjian and
                  Daneshzand, Mohammad and
                  Gensollen, Nicolas and
                  Proulx, Nicole and
                  Focke, Niels and
                  Chalas, Nikolas and
                  Shubi, Omer and
                  Mainar, Pablo and
                  Sundaram, Padma and
                  Silva, Pedro and
                  Molfese, Peter J and
                  Das, Proloy and
                  Chu, Qian and
                  Li, Quanliang and
                  Barthélemy, Quentin and
                  Nadkarni, Rahul and
                  Gatti, Ramiro and
                  Apariciogarcia, Ramonapariciog and
                  Aagaard, Rasmus and
                  Nasri, Reza and
                  Koehler, Richard and
                  Stargardsky, Riessarius and
                  Oostenveld, Robert and
                  Seymour, Robert and
                  Schirrmeister, Robin Tibor and
                  Law, Ryan and
                  Pai, Sagun and
                  Perry, Sam and
                  Louviot, Samuel and
                  Saha, Sawradip and
                  Mathot, Sebastiaan and
                  Major, Sebastian and
                  Treguer, Sebastien and
                  Castaño, Sebastián and
                  Deng, Senwen and
                  Antopolskiy, Sergey and
                  Wong, Simeon and
                  Wong, Simeon and
                  Poil, Simon-Shlomo and
                  Foslien, Sondre and
                  Singh, Sourav and
                  Chambon, Stanislas and
                  Bethard, Steven and
                  Gutstein, Steven M and
                  Meyer, Svea Marie and
                  Wang, T and
                  Donoghue, Thomas and
                  Moreau, Thomas and
                  Radman, Thomas and
                  Gates, Timothy and
                  Ma, Tom and
                  Stone, Tom and
                  Clausner, Tommy and
                  Anijärv, Toomas Erik and
                  Xia, Xiaokai and
                  Zuo, Yiping and
                  Zhang, Zhi and
                  buildqa and
                  luzpaz},
  title        = {MNE-Python},
  month        = jan,
  year         = 2024,
  publisher    = {Zenodo},
  version      = {v1.6.1},
  doi          = {10.5281/zenodo.10519948},
  url          = {https://doi.org/10.5281/zenodo.10519948},
}

@article{logothetis2008we,
  title={What we can do and what we cannot do with fMRI},
  author={Logothetis, Nikos K},
  journal={Nature},
  volume={453},
  number={7197},
  pages={869--878},
  year={2008},
  publisher={Nature Publishing Group UK London}
}

@article{king2014characterizing,
  title={Characterizing the dynamics of mental representations: the temporal generalization method},
  author={King, Jean-R{\'e}mi and Dehaene, Stanislas},
  journal={Trends in cognitive sciences},
  volume={18},
  number={4},
  pages={203--210},
  year={2014},
  publisher={Elsevier}
}

@article{schonmann2025stimulus,
  title={Stimulus dependencies—rather than next-word prediction—can explain pre-onset brain encoding during natural listening},
  author={Sch{\"o}nmann, In{\'e}s and Szewczyk, Jakub and de Lange, Floris P and Heilbron, Micha},
  journal={bioRxiv},
  pages={2025--03},
  year={2025},
  publisher={Cold Spring Harbor Laboratory}
}

@article{zada2025podcast,
  title={The “Podcast” ECoG dataset for modeling neural activity during natural language comprehension},
  author={Zada, Zaid and Nastase, Samuel A and Aubrey, Bobbi and Jalon, Itamar and Michelmann, Sebastian and Wang, Haocheng and Hasenfratz, Liat and Doyle, Werner and Friedman, Daniel and Dugan, Patricia and others},
  journal={Scientific Data},
  volume={12},
  number={1},
  pages={1135},
  year={2025},
  publisher={Nature Publishing Group UK London}
}

@article{gwilliams2025hierarchical,
  title={Hierarchical dynamic coding coordinates speech comprehension in the human brain},
  author={Gwilliams, Laura and Marantz, Alec and Poeppel, David and King, Jean-Remi},
  journal={bioRxiv},
  pages={2024--04},
  year={2025}
}

@article{carlson2013representational,
  title={Representational dynamics of object vision: the first 1000 ms},
  author={Carlson, Thomas and Tovar, David A and Alink, Arjen and Kriegeskorte, Nikolaus},
  journal={Journal of vision},
  volume={13},
  number={10},
  pages={1--1},
  year={2013},
  publisher={The Association for Research in Vision and Ophthalmology}
}

@article{king2014two,
  title={Two distinct dynamic modes subtend the detection of unexpected sounds},
  author={King, Jean-R{\'e}mi and Gramfort, Alexandre and Schurger, Aaron and Naccache, Lionel and Dehaene, Stanislas},
  journal={PloS one},
  volume={9},
  number={1},
  pages={e85791},
  year={2014},
  publisher={Public Library of Science San Francisco, USA}
}

\newpage
\clearpage
\section{Supplementary Information}
\setcounter{figure}{0} 
\renewcommand{\thefigure}{S\arabic{figure}}

\begin{figure}[htbp]
    \centering
    \includegraphics[width=1\linewidth]{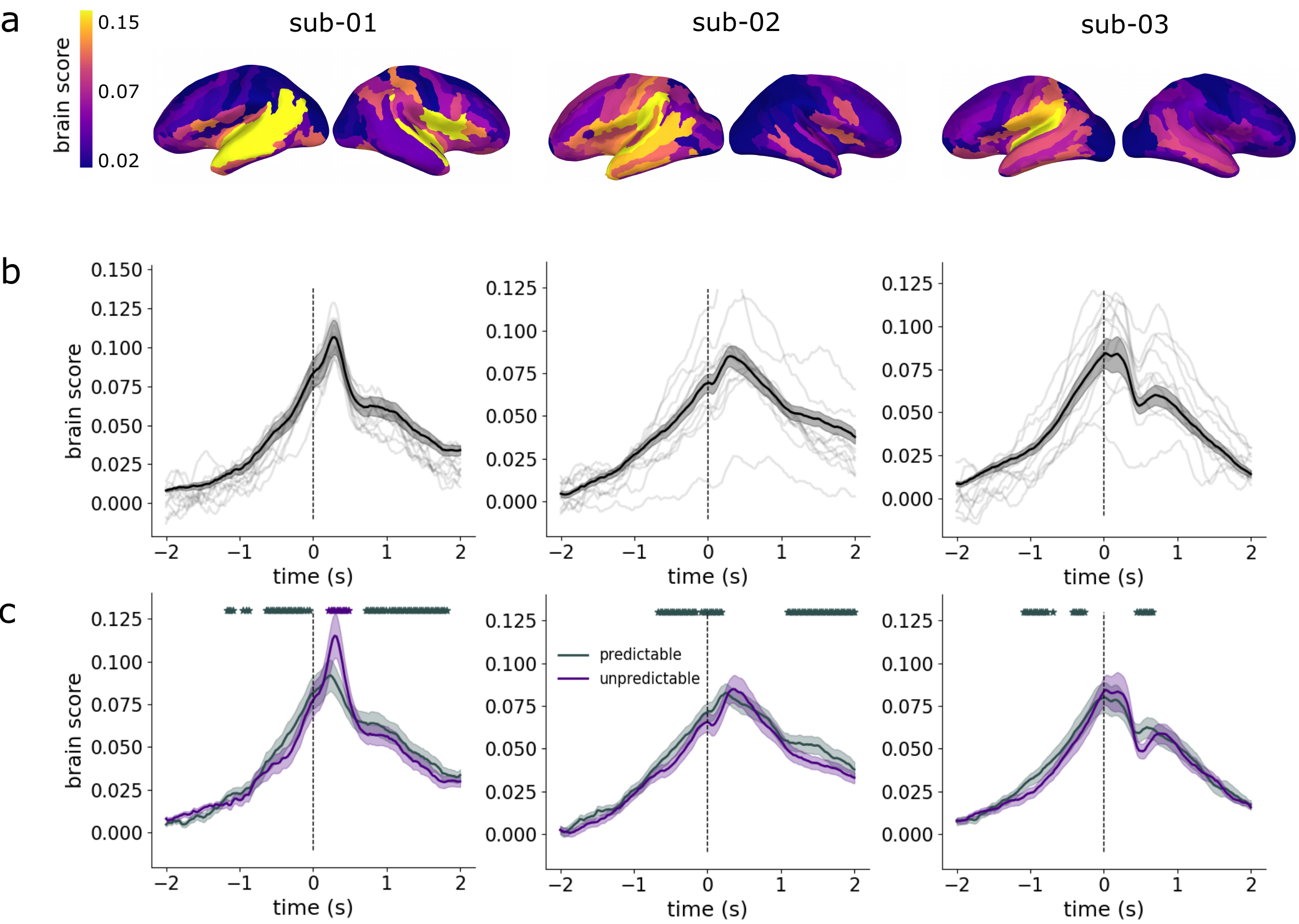}
    \caption{\textbf{Encoding results for individual subjects in the MEG dataset} To examine whether predictable words are better encoded before word onset, we categorized words as predictable or unpredictable based on whether they appeared in the top-5 predictions of GPT-2 for that position \citep{goldstein2022}. 
    \textbf{a.} Brain score values peak in the left hemisphere, especially in the temporal cortex and inferior frontal areas associated with language processing.
    \textbf{b.} Brain scores from single (gray) and all 10 (black) sessions show encoding before word onset, peaking at or shortly after it (see Methods for MEG source selection).  
    \textbf{c.} The encoding model was trained separately to compute brain scores for predictable and unpredictable words.
    The figure indicates that predictable words show stronger encoding up to 1 second before word onset. The encoding of unpredictable words appears stronger around 400 ms post-onset, though these effects did not reach statistical significance for subjects 2 and 3. Star symbols mark significant differences between predictable and unpredictable words calculated using a dependent t-test for paired samples on the brain scores of the two groups across the same MEG sources and accounted for multiple hypothesis testing using the Benjamini-Hochberg correction. we considered $q$ values smaller than $0.05$ as significant.  A smoothing window of 200ms was used in this analysis.  All error bars were computed as the standard error across the MEG sources. }
    \label{fig:supp_MEG_fig2_per_subj}
\end{figure}

\newpage

\begin{figure}[htbp]
    \centering
    \includegraphics[width=0.8\linewidth]{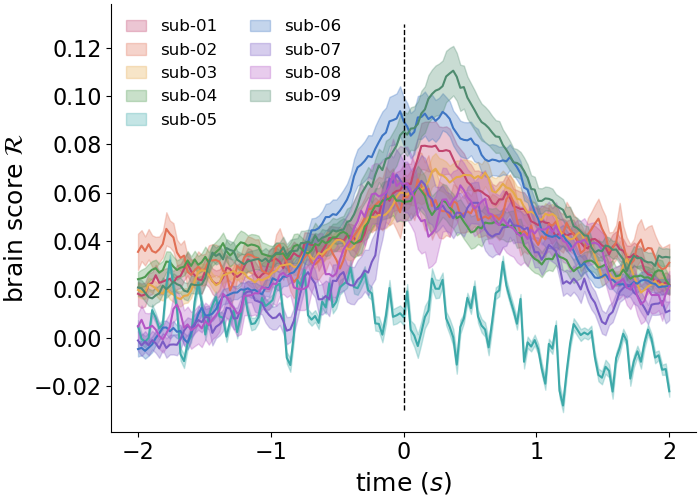}
    \caption{\textbf{Encoding values for individual subjects in the ECoG dataset.}
    Brain score values and temporal profiles vary across subjects in the ECoG dataset, partly due to differences in the number and locations of electrodes selected for each participant. Shaded regions in the line plots indicate the standard error of the mean (SEM) across ECoG electrodes included for each subject.}
    
    \label{fig:supp_ECoG_fig2_per_subj}
\end{figure}

\newpage

\begin{figure}[!ht]
    \centering
    \includegraphics[width=0.8\linewidth]{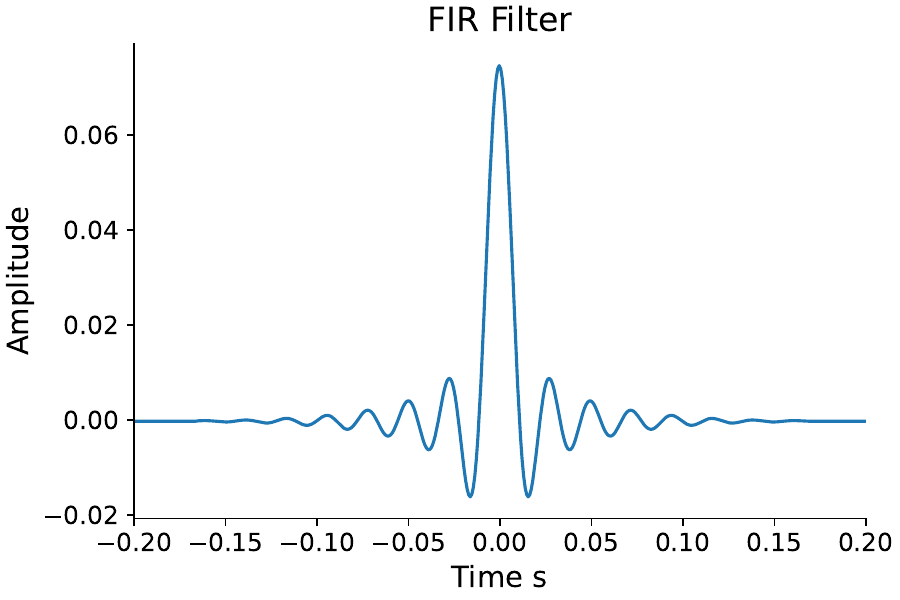}
    \caption{\textbf{Shape of the FIR filter used for band-pass filtering MEG data between 0.1–40 Hz.}
    Shown is the finite impulse response (FIR) filter implemented in MNE and used for MEG preprocessing. Because this filter is non-causal and applied in both forward and reverse directions, it can, in principle, introduce temporal leakage. However, the filter amplitude converges to zero within the first 100 ms, making any temporal leakage negligible for the effects examined in this study.}
    \label{fig:FIR_filter}
\end{figure}

\begin{figure}[!ht]
    \centering
    \includegraphics[width=1\linewidth]{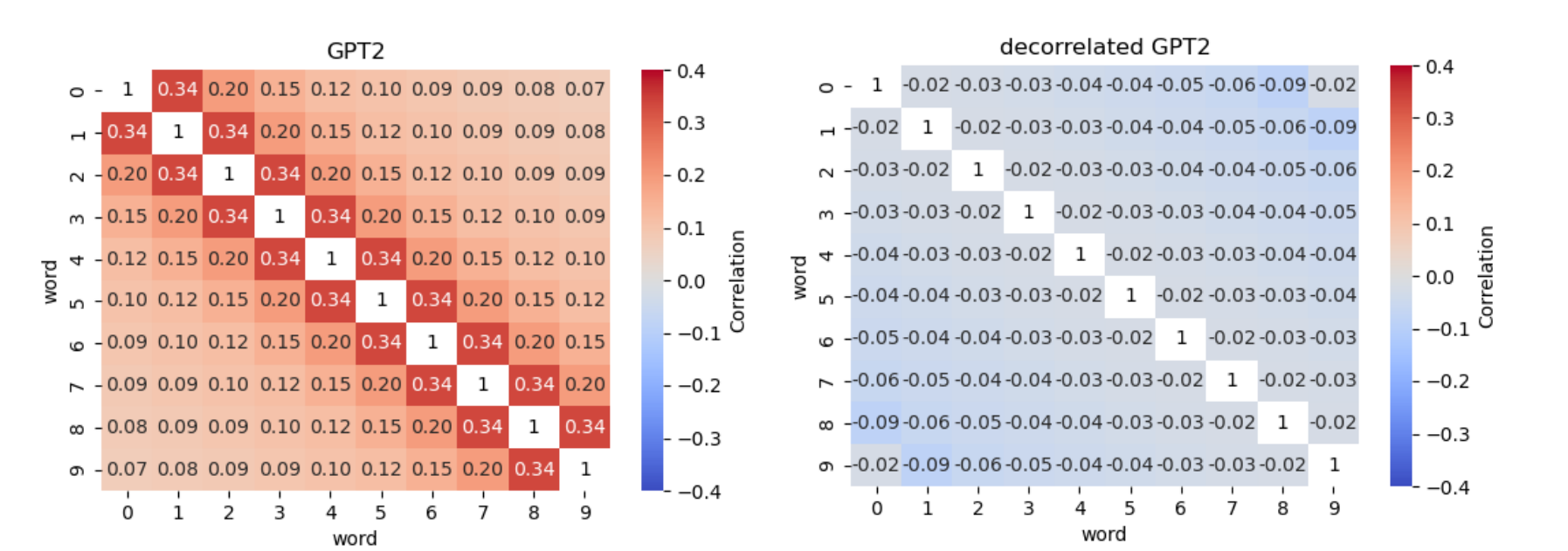}
    \caption{\textbf{Average correlation between nearby word embeddings in the 10h MEG dataset.} 
    (left) GPT2 embeddings have large, symmetric correlations between nearby words. (right) After removing the correlation of each word embedding in the story with its previous 8 word embeddings, most of the correlations between nearby word embeddings disappear. These figures were obtained by calculating the correlation matrix given each 10 consecutive words (in non-overlapping windows), and then averaging all those values across the 10h MEG dataset.
    }
    \label{fig:supp_embedding_correlations}
\end{figure}

\newpage

\begin{figure}[!ht]
    \hspace{-1cm} 
    \centering
    \includegraphics[width=1\textwidth]{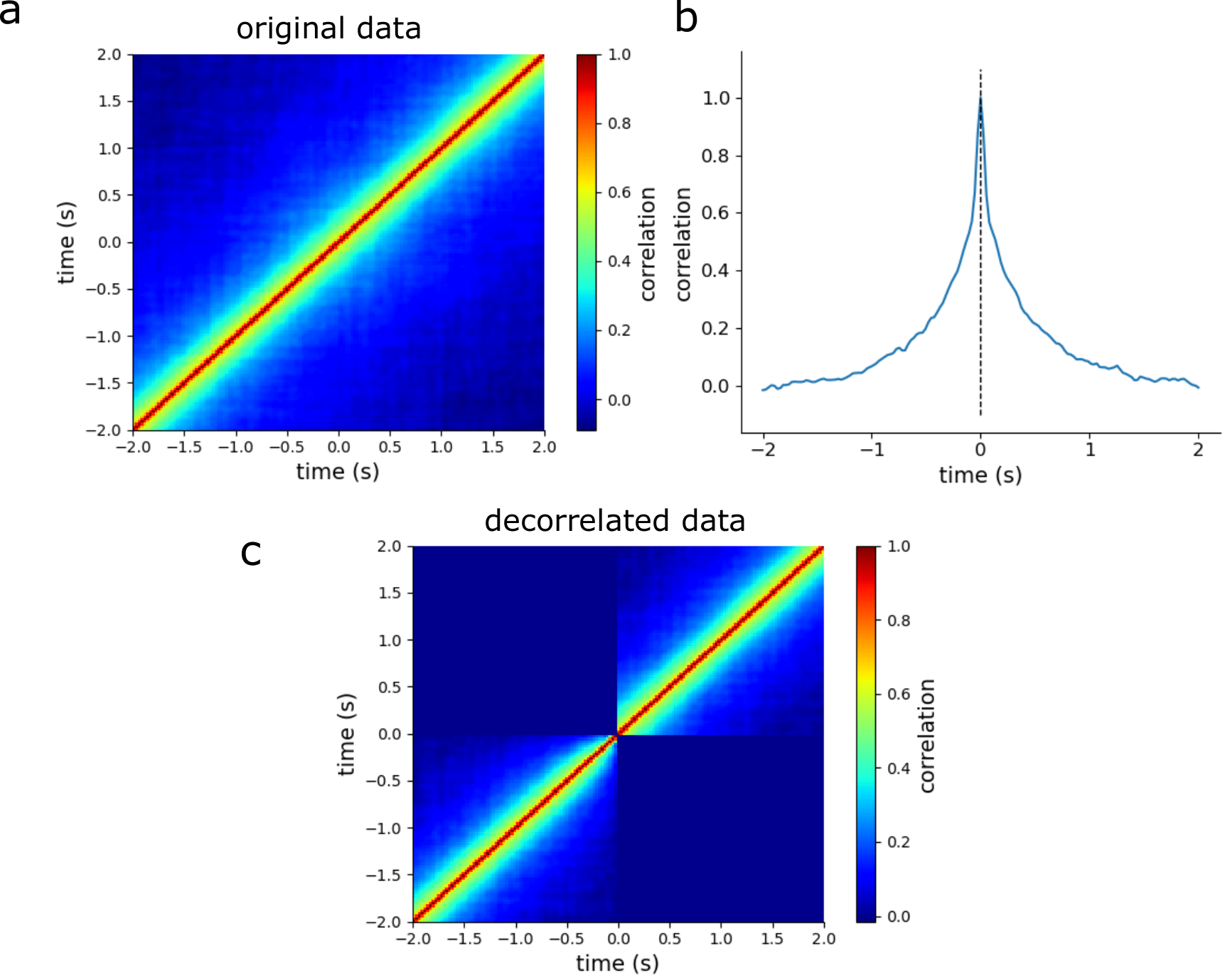}
    \caption{\textbf{Temporal correlation within the epochs in the MEG dataset.}
    \textbf{a.} shows the time-by-time correlation matrix of the MEG signal across the epochs. 
    \textbf{b.} shows a slice at t=0 of the correlation matrix. The autocorrelation decays rapidly and reaches zero within $\sim$\,1 second.
    \textbf{c} shows the correlation matrix after regressing out all post-onset time-points from each pre-onset time-point.
    }
    
    \label{fig:supp_temporal_correlation}
\end{figure}

\newpage
The temporal generalization (TG) matrix revealed negative generalization areas approximately one second from the diagonal (\fig\ref{fig:fig_TG} and  \fig\ref{fig:fig_TG_control}), suggesting an inversion of the neural code over time. Similar negative–positive patterns have been reported in MEG studies of visual and auditory processing \citep{carlson2013representational, king2014two, king2014characterizing} and are often attributed to reversals in neural activity patterns.

In our case, because the encoding model is linear and maps embeddings to individual MEG time points, the negative generalization might reflect a sign flip in the underlying neural response from pre- to post-onset, potentially arising from the oscillatory nature of the MEG signal. Although we do not observe such a sign flip in the correlation matrix computed across the full dataset (\fig\ref{fig:supp_temporal_correlation}), it could still exist within the subset of words that most strongly drive the encoding model.

\begin{figure}[htpb]
    \hspace{-1cm} 
    \centering
    \includegraphics[width=1\textwidth]{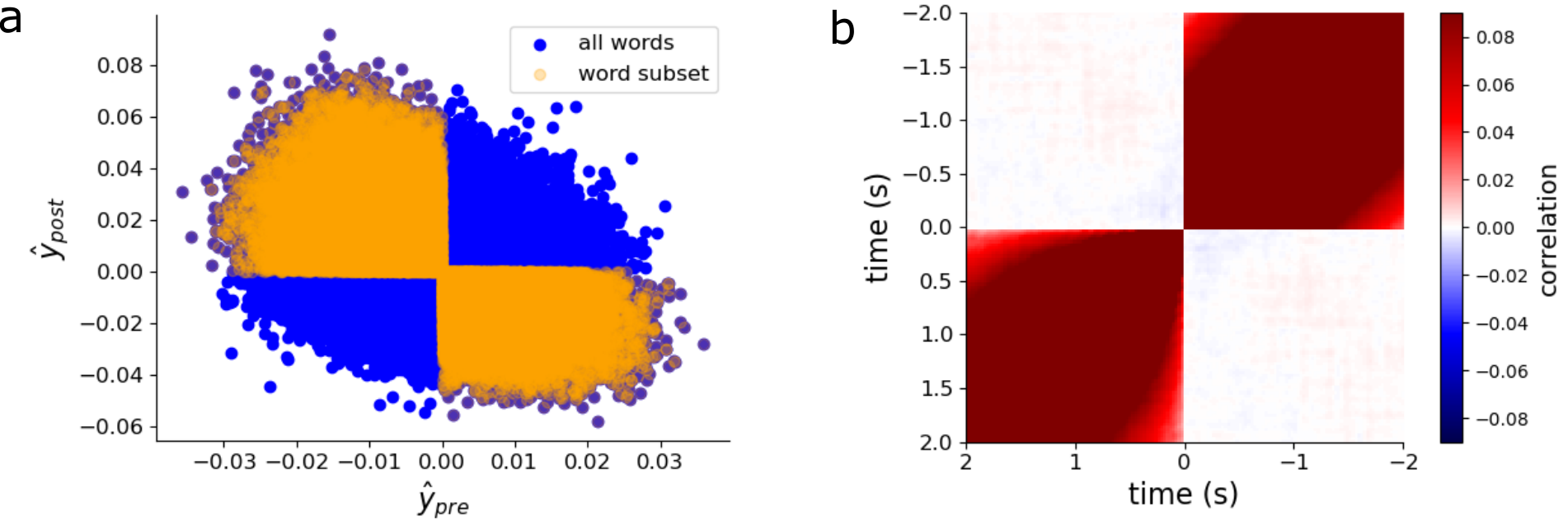}
    \caption{\textbf{Negative areas in the TG matrix are not caused by a sign flip in brain activity.}
    \textbf{a.} Average model predictions for pre- and post-onset activity (averaged over [-600, 0] ms and [0, 600] ms, respectively). For each MEG source, a subset of words showing anticorrelated model predictions is highlighted in orange. This analysis was performed on data from the first participant in the MEG dataset after decorrelating pre- and post-onset signals, similar to \fig\ref{fig:fig_TG_control}.
    \textbf{b.} Correlation matrix computed for the selected subset of words at each MEG source and averaged across the 30 MEG sources used in the main analyses. The resulting matrix shows that even for this subset, there is no systematic sign reversal in brain activity before and after word onset; instead, the two signals remain largely uncorrelated.}
    \label{fig:supp_correlation_word_subset}
\end{figure}

\begin{figure}[htpb]
    \centering
    \includegraphics[width=1\linewidth]{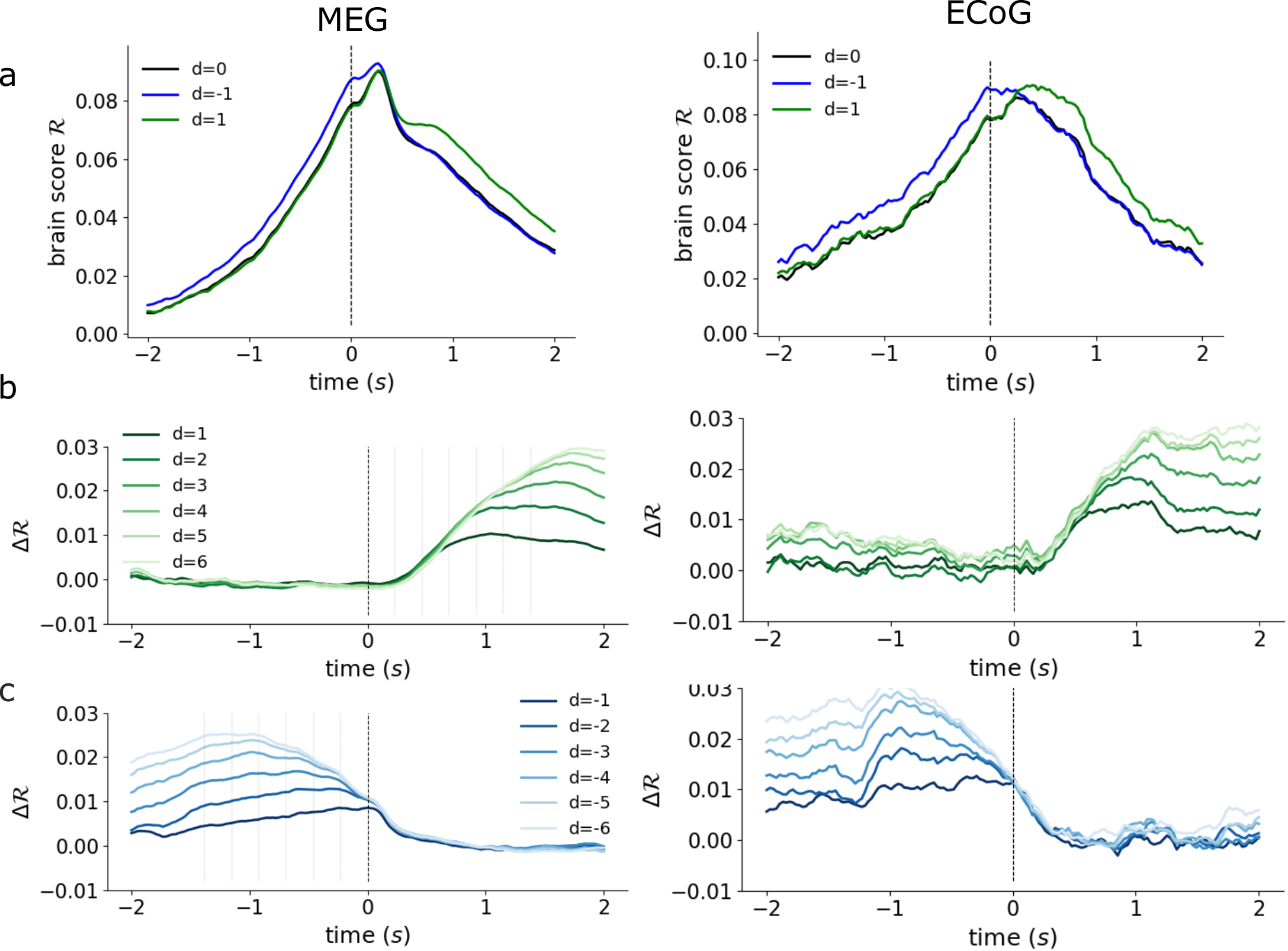}
    \caption{\textbf{Analysis with original gpt2 embeddings also shows that future word embeddings do not improve encoding, while past word embeddings do.}
    This figure shows was obtained similar to \fig\ref{fig:forecast} but using original gpt2 embeddings, instead of decorrelated embeddings.
    All curves represent averages across participants.
    The embedding vector is constructed by concatenating $d$ future word embeddings ($d > 0$) or $|d|$ past word embeddings ($d < 0$) along with the embedding of the current word $w_i$.
    \textbf{a} Including the next word embedding in the encoding model ($d=1$) enhances encoding only after that word is heard in the story, while including the previous word ($d=-1$) improves encoding even after the current word’s onset.
     Encoding enhancement, $\Delta\mathcal{R}$, is shown for \textbf{b} negative and \textbf{c} positive values of $d$. Vertical gray lines mark the median inter-word interval values. Adding each successive future word embedding improves encoding only after that word is heard in the narrative, while including previous words consistently improves encoding beyond their offset.
    }
    
    \label{fig:forecast_original}
\end{figure}

\begin{figure}[htpb]
    \centering
    \includegraphics[width=0.7\linewidth]{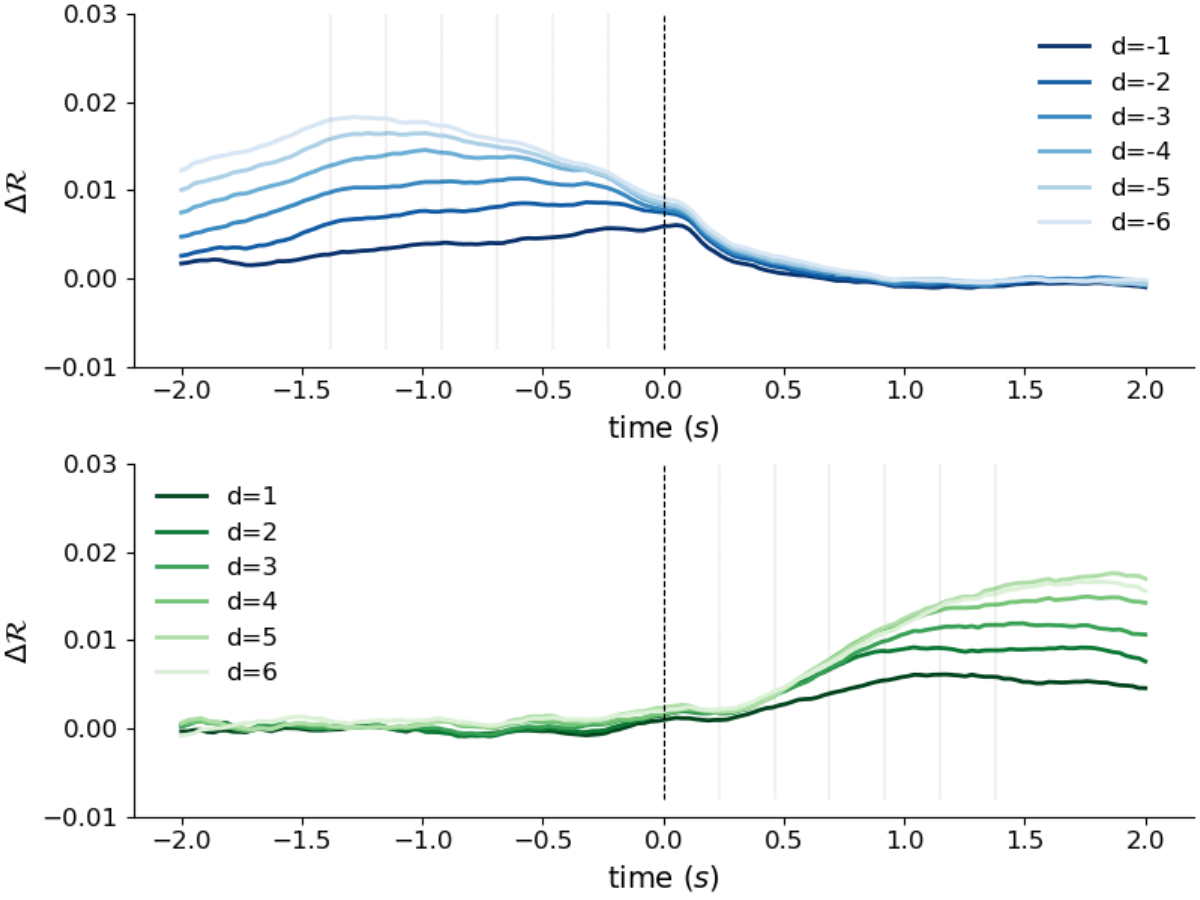}
    \caption{\textbf{Lack of evidence for future-word encoding is not due to MEG source selection.}
    To test whether the absence of an effect observed in \fig\ref{fig:forecast} could be explained by our MEG source selection procedure, we repeated the analysis using only sources that, on average, showed enhanced encoding within the [0, 230] ms window for $d > 0$. Even with this intentionally biased selection, encoding did not noticeably improve within the [0, 230] ms interval. In contrast, including embeddings of previous words resulted in a measurable increase in encoding performance.}
    \label{fig:forecast_different_regions}
\end{figure}

\begin{figure}[htpb]
    \hspace{-1cm} 
    \centering
    \includegraphics[width=1\textwidth]{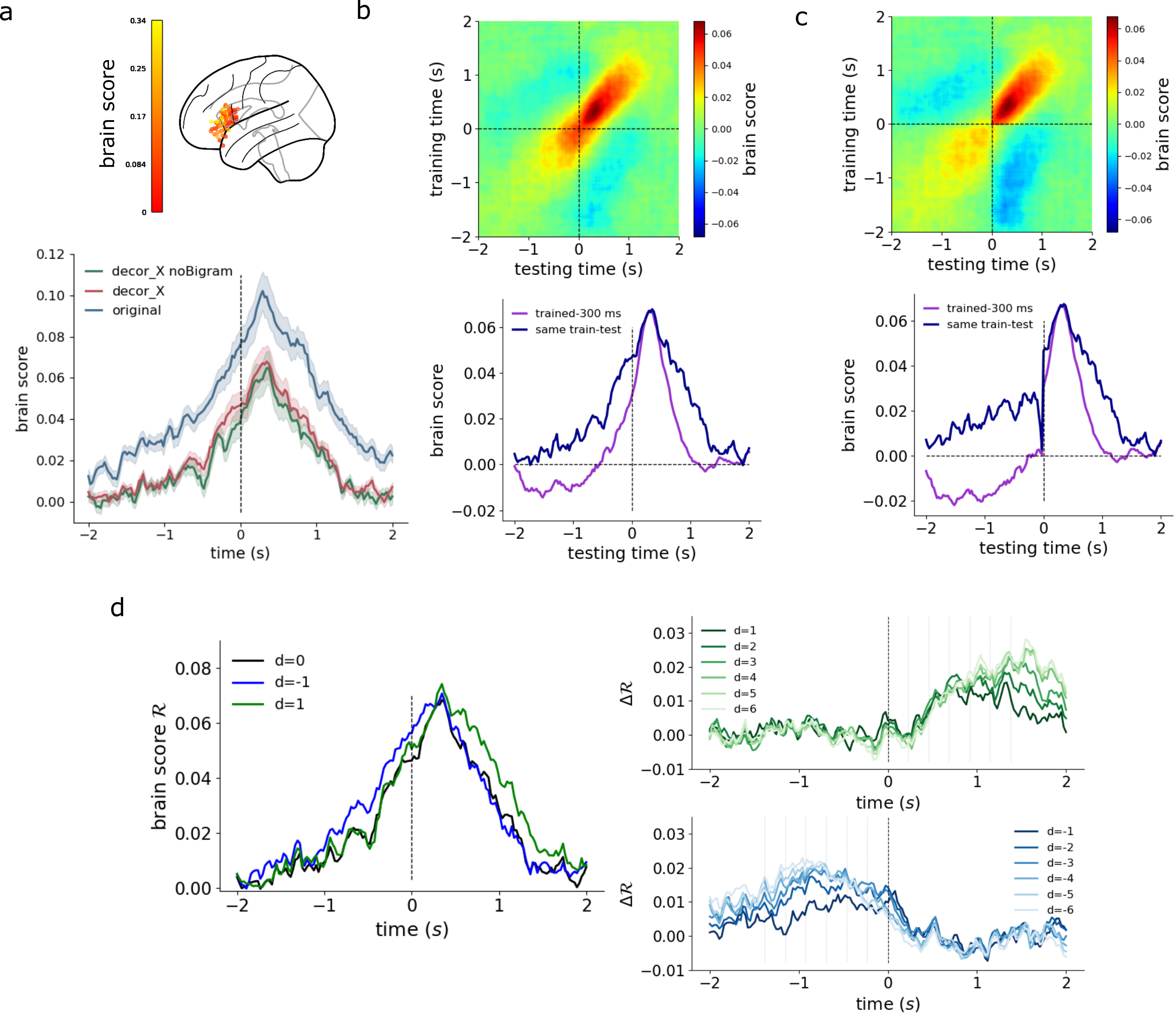}
    \caption{\textbf{No evidence for predictive pre-activation in the IFG region of the ECoG dataset.}
    To test whether the absence of effects was due to the broad selection of ECoG electrodes, we repeated the analysis using only electrodes located in the inferior frontal gyrus (IFG), previously reported to show robust pre-onset encoding of upcoming stimuli \citep{goldstein2022}. The same qualitative patterns were observed in the IFG as in the borader selection of electrodes in the main text.\textbf{a.} \textbf{top}: location of the IFG electrodes included in this analysis.\textbf{bottom}: A comparison of encoding models shows that removing correlations between neighboring word embeddings (decor\_X) doesn't eliminate pre-onset encoding. Similarly, eliminating bi-grams in the narrative (decor\_X noBigram) slightly reduces overall encoding performance, but the pre-onset effect persists.
    The shaded regions in the line plots indicate the standard error of the mean (SEM) across aggregated ECoG electrodes/MEG sources of all participants. \textbf{b.} Temporal generalization of representations captured by the encoding model differs before and after word onset. \textbf{top}: Temporal generalization (TG) matrix computed with decorrelated embeddings . Positive values indicate successful generalization of representations across time. 
    \textbf{bottom}: Generalization profiles for models trained at the peak encoding response ($\sim$\,300 ms; purple) and along the diagonal (blue). The divergence between the two curves indicates that pre-onset encoding does not reflect the same representations engaged during word processing., \textbf{c.} Removing autocorrelation between pre- and post-onset activity does not eliminate pre-onset encoding.
    \textbf{top} Temporal generalization (TG) matrix computed using decorrelated embeddings after regressing out post-onset brain activity from the pre-onset signal. Removing correlations in the neural signal should, in principle, eliminate any trace of predictive pre-activation. Following this procedure, the small pre-onset generalization observed near word onset in panel b. disappears.
    \textbf{bottom} Generalization profiles for models trained at the peak encoding response ($\sim$\,300 ms; purple) and along the diagonal (blue). The persistence of pre-onset encoding (blue curve) despite the absence of pre-activation indicates that pre-onset encoding is not necessarily a signature of prediction. .\textbf{d.} Encoding of the future and past words.
    All curves represent averages across participants.
    The embedding vector is constructed by concatenating $d$ future word embeddings ($d > 0$) or $|d|$ past word embeddings ($d < 0$) along with the embedding of the current word $w_i$.
    \textbf{left} Including the next word embedding in the encoding model ($d=1$) enhances encoding only after that word is heard in the story, while including the previous word ($d=-1$) improves encoding even after the current word’s onset.
     Encoding enhancement, $\Delta\mathcal{R}$, is shown for \textbf{top right} positive and \textbf{bottom right} positive values of $d$. Vertical gray lines mark the median inter-word interval values. Adding each successive future word embedding improves encoding only after that word is heard in the narrative, while including previous words consistently improves encoding beyond their offset.}
    \label{fig:supp_IFG}
\end{figure}

\begin{figure}[htpb]
    \centering
    \includegraphics[width=1\linewidth]{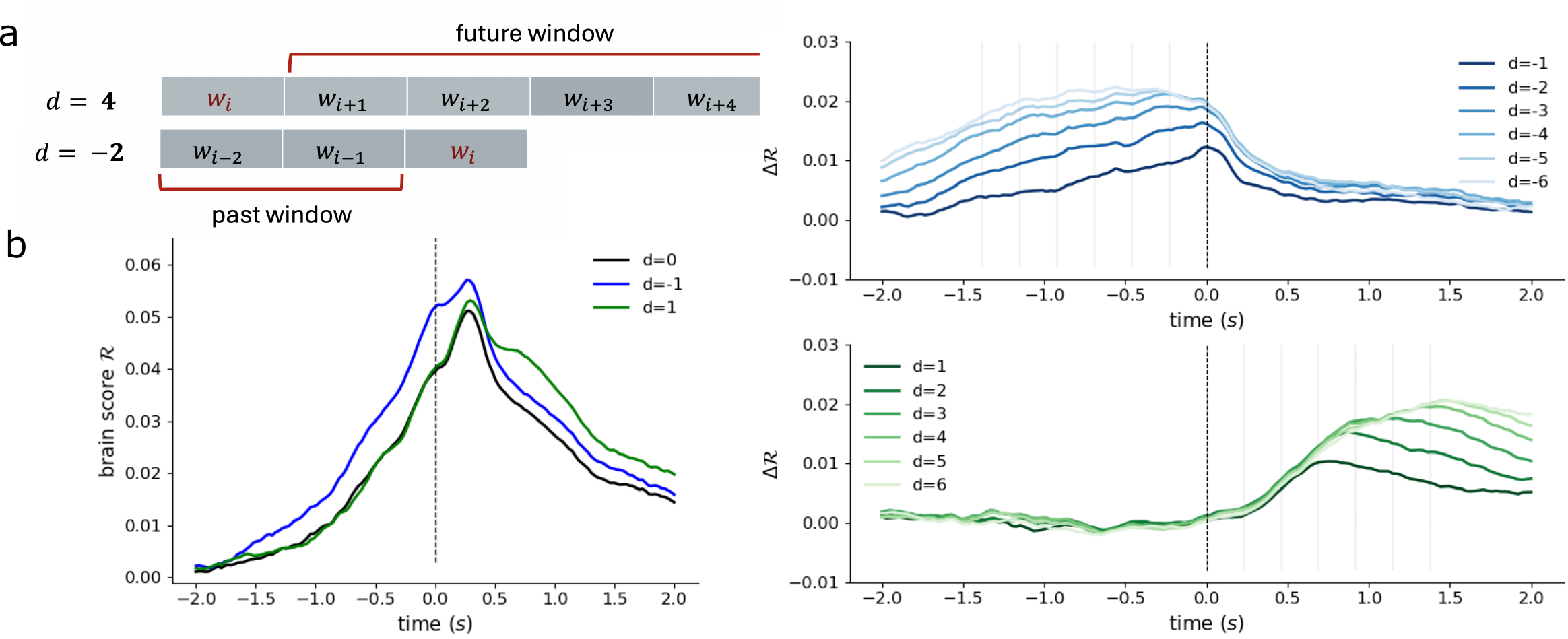}
    \caption{\textbf{No evidence for future-word encoding replicated using GloVe embeddings in MEG data.}
    To test whether the absence of future-word encoding observed in \fig\ref{fig:forecast} could be attributed to the contextual nature of GPT embeddings, we repeated the analysis using static GloVe embeddings. 
    \textbf{a.} The variable $d$ represents the number of word embeddings concatenated with the current word embedding to form the final vector. Positive values of $d$ signify the inclusion of future word embeddings and negative $d$ signify the inclusion of past word embeddings.
    \textbf{b. and c.} The same qualitative pattern was observed: including future-word embeddings did not enhance model performance within the [0, 230] ms window after word onset, whereas including past-word embeddings clearly improved it.}
    \label{fig:forecast_glove}
\end{figure}

To test this, we examined encoding model predictions for each word in the story and selected those for which the model predicts a sign reversal across word onset (\fig\ref{fig:supp_correlation_word_subset}.a). If the negative TG values were caused by a true sign flip in the MEG signal, this subset should exhibit negative correlations between pre- and post-onset neural responses. However, \fig\ref{fig:supp_correlation_word_subset}.b shows that the recorded MEG activity for these words does not switch sign across onset.

Together, these findings suggest that the negative values in the TG analysis are not simply driven by a corresponding reversal in the neural signal itself. In fact, \citet{gwilliams2025hierarchical} showed in a simulation that this type of negative–positive diagonal pattern can emerge from a spatio-temporal dynamic neural code.  They suggest that the neural representation of an item in a temporal sequence , such as a word in a sentence, moves through neural representational space over time. This effectively creates room for the next item to be encoded while still maintaining the earlier item in memory, avoiding interference between them.\\

\end{document}